\documentclass[aps,pra,twocolumn,nofootinbib]{revtex4}

\usepackage[english]{babel}
\usepackage{amsmath}
\usepackage{amsthm}
\usepackage{amssymb}
\usepackage{booktabs}
\usepackage{color}
\usepackage{hyperref}
\hypersetup{
  colorlinks   = true,  %Colours links instead of ugly boxes
  urlcolor     = blue,  %Colour for external hyperlinks
  linkcolor    = blue,  %Colour of internal links
  citecolor    = red    %Colour of citations
}
\usepackage{natbib}
\usepackage{pgfplots}
\usepackage{subfigure}
\usepackage{url}

\newcommand{\refFigure}[1]{{\textrm{Fig.~\ref{#1}}}}

\newcommand{\refEquation}[1]{{\textrm{Eq.~(\ref{#1})}}}

\usepackage{scalerel,stackengine}
\stackMath
\newcommand\reallywidehat[1]{%
\savestack{\tmpbox}{\stretchto{%
  \scaleto{%
    \scalerel*[\widthof{\ensuremath{#1}}]{\kern-.6pt\bigwedge\kern-.6pt}%
    {\rule[-\textheight/2]{1ex}{\textheight}}%WIDTH-LIMITED BIG WEDGE
  }{\textheight}% 
}{0.5ex}}%
\stackon[1pt]{#1}{\tmpbox}%
}
\usepackage{comment}
\usepackage{float}
\usepackage{bm}

\begin{document}

\title{Van der Waals universality near a quantum tricritical point}

\author{P. M. A. \surname{Mestrom}}
%\email{p.m.a.mestrom@tue.nl}
\altaffiliation[Corresponding author: ]{p.m.a.mestrom@tue.nl}

\author{V. E. \surname{Colussi}}

\author{T. \surname{Secker}}

\author{G. P. \surname{Groeneveld}}

\author{S. J. J. M. F. \surname{Kokkelmans}}
\affiliation{Eindhoven University of Technology, P.~O.~Box 513, 5600 MB Eindhoven, The Netherlands}

\date{\today}

%\pacs{31.15.-p, 34.50.-s, 67.85.-d \del{??????????}}

\begin{abstract}
We study the three-body scattering hypervolume $D$ of atoms whose scattering length $a$ is on the order of or smaller than the typical range $r_{\mathrm{vdW}}$ of the van der Waals attraction. We find that the real part of $D$ behaves universally in this weakly interacting regime ($|a|/r_{\mathrm{vdW}}\lesssim 1$) in the absence of trimer resonances. This universality originates from hard-spherelike collisions that dominate elastic three-body scattering. We use this result to make quantitative predictions for the thermodynamics and elementary excitations of an atomic Bose-Einstein condensate in the vicinity of a quantum tricritical point, including quantum droplets stabilized by effective three-body interactions. 
\end{abstract}

\maketitle

\textit{Introduction.}---The thermodynamics of ultracold dilute quantum gases is determined by the asymptotic behavior of few-body scattering processes. Bosonic many-body systems can be universally described with only one parameter to characterize the two-body interactions: the $s$-wave scattering length $a$ \cite{huang1963statistical}. However, even in the limit $a \to 0$ when two-body interactions effectively vanish, system properties are still determined by few-body scattering. The quantum phase diagram of a uniform Bose-Einstein condensate (BEC) in this limit is determined by the three-body scattering hypervolume $D$ that is the analog of $a$ \cite{tan2008hypervolume,zwerger2019phasetransition}. Provided $\mathrm{Re}(D)>0$, a quantum tricritical point exists at $a = 0$ where the boundaries between liquid and gaseous ground states and vacuum meet \cite{zwerger2019phasetransition}. Quantum tricritical points are rare, occuring for certain metallic magnets \cite{belitz2017tricritical, friedemann2018tricritical, canfield2018tricritical} and for a generalized Dicke model \cite{xu2019tricritical}.

%\textit{Introduction.}---The thermodynamics of ultracold dilute quantum gases is \ins{strongly influenced by interatomic forces. The few-body scattering processes resulting from these interactions determine the properties of BECs in a hierarchic way. The dominant process is two-body scattering which can be predominantly characterized by the $s$-wave scattering length $a$ to describe BECs. However, in the limit \ins{$a \to 0$} when two-body interactions effectively vanish, system properties are significantly influenced by three-body scattering processes which are encapsulated by the three-body scattering hypervolume $D$ that is the analog of $a$ \cite{tan2008hypervolume}.} 
%\ins{Consequently, $D$ determines the quantum phase diagram of a BEC near $a = 0$ \cite{zwerger2019phasetransition}. If $\mathrm{Re}(D)>0$, there exists a first-order phase transition from the vacuum to a liquid state for $a<0$. For $a>0$, there is a second-order phase transition from the vacuum to a gaseous state. The corresponding boundaries meet at the quantum tricritical point $a = 0$ \cite{zwerger2019phasetransition}. Although quantum tricritical points are very rare, they also occur in the phase diagram of certain metallic magnets \cite{belitz2017tricritical, friedemann2018tricritical, canfield2018tricritical} and a generalized Dicke model \cite{xu2019tricritical}.}

%\del{maybe add Helium to intro like Zwerger (2019)}

The ground-state energy density $\mathcal{E}$ of a uniform BEC at $a = 0$ is determined by $D$ via
\begin{equation}
\mathcal{E}(n) = \frac{\hbar^2 D n^3}{6 m} +  ...
\end{equation}
where the dots indicate terms with higher powers of the number density $n$ \cite{tan2008hypervolume}. In the liquid phase when $a<0$ and $\mathrm{Re}(D)>0$, effective three-body repulsion and two-body attraction compete. This provides a three-body stabilization mechanism for liquid quantum droplets against collapse \cite{bulgac2002droplets}. In Ref.~\cite{gao2004dropletsvdW}, it was suggested that the experimentally observed collapse \cite{gerton2000collapseBEC, roberts2001collapseBEC, donley2001collapseBEC, eigen2016collapseBEC} is a relaxation process towards the liquid equilibrium state. Recently, quantum droplets were predicted \cite{petrov2015dropletsBoseBoseMix} and experimentally observed in both mixed \cite{cabrera2018quantumdroplet, tarruell2018quantumdroplet, semeghini2018quantumdroplet} and dipolar BECs \cite{kadau2016dipolarquantumdroplet, ferrierbarbut2016dipolarquantumdroplet, chomaz2016dipolarquantumdroplet}, however, these systems were based on two-body stabilization mechanisms. Additionally, in typical alkali systems, $D$ acquires an imaginary part proportional to the three-body recombination rate \cite{tan2017hypervolume,braaten2006universality}, so that these states are inherently metastable. Ultimately, a quantitative understanding of droplet properties near the quantum tricritical point depends on the sign and magnitude of $\mathrm{Re}(D)$ for realistic systems.
% \ins{which we determine in this Letter}.
 
For weakly interacting systems, Refs.~\cite{tan2008hypervolume, tan2017hypervolume, mestrom2019hypervolumeSqW, zwerger2019phasetransition} investigated the scattering hypervolume, demonstrating how $D$ is influenced by nonuniversal three-body quasibound states \cite{tan2017hypervolume, mestrom2019hypervolumeSqW} and is connected to physical observables at $a = 0$ \cite{zwerger2019phasetransition}. 
%\ins{Nevertheless, a study of $D$ involving two-body interaction models that contain the long-range atomic van der Waals attraction is still missing.}
% \del{Nevertheless, none of these studies involved two-body interaction models that contain the long-range atomic van der Waals attraction.}
Nevertheless, none of these studies solved the three-body problem for two-body interaction models that contain the long-range atomic van der Waals attraction.
We find that this is the essential ingredient for quantitative predictions of $D$, whose real part is universally fixed by the van der Waals range $r_{\mathrm{vdW}} = (m C_6/\hbar^2)^{1/4}/2$, where $C_6$ is the dispersion coefficient describing the long-range behavior of the interatomic interaction. 
Although the mechanisms differ, van der Waals universality also determines $D$ in the strongly interacting regime $(|a|/r_{\mathrm{vdW}} \gg 1)$ by setting the spectrum of Efimov trimers \cite{berninger2011cesium133, naidon2017review, greene2017review, dincao2018review, wang2012origin, dincao2012heteronuclear, schmidt2012universality, naidon2014physicalorigin, blume2015TBPhelium, naidon2012threebodyparameter4He, sorensen2013recombinationoptical, chapurin2019precision}.

%$\mathrm{Re}(D)$ was first studied for strongly interacting systems in the context of Efimov physics \cite{efimov1979threebody, braaten2002diluteBEC, braaten2006universality, dincao2018review, mestrom2019hypervolumeSqW}. The corresponding universal relations were numerically studied for contact interactions \cite{braaten2002diluteBEC} and only recently for local finite-range potentials \cite{mestrom2019hypervolumeSqW}. For weakly interacting systems, only a few studies \cite{tan2008hypervolume, tan2017hypervolume, mestrom2019hypervolumeSqW, zwerger2019phasetransition} investigated the scattering hypervolume, demonstrating how $D$ is influenced by nonuniversal three-body quasibound states \cite{tan2017hypervolume, mestrom2019hypervolumeSqW} and how it is connected to physical observables at $a = 0$ \cite{zwerger2019phasetransition}. Nevertheless, none of these studies involved two-body interaction models that contain the van der Waals attraction. This is an essential ingredient for quantitative predictions of $D$ for alkali atoms.

In this Letter, we present a numerical study of the scattering hypervolume for identical bosons interacting via pairwise van der Waals potentials in the weakly interacting regime ($|a|\lesssim r_{\mathrm{vdW}}$). We find that $\mathrm{Re}(D)$ is predominantly determined by $a$ and $r_{\mathrm{vdW}}$ and analyze the origin of this van der Waals universality. This is used to make universal quantitative predictions for atomic BECs near the quantum tricritical point, including quantum droplets stabilized by effective three-body interactions.

\textit{Method.}---Here we use the Alt, Grassberger, and Sandhas (AGS) approach \cite{alt1967ags} which has been proven to be a powerful method for calculating the scattering hypervolume \cite{mestrom2019hypervolumeSqW}. The AGS equations,
\begin{equation}
\begin{aligned}
U_{0 0}(z) &= \sum_{\alpha = 1}^{3} T_{\alpha}(z) G_0(z) U_{\alpha 0}(z), \\
U_{\alpha 0}(z) &= G_0^{-1}(z) + \sum_{\substack{\beta = 1 \\ \beta \neq \alpha}}^{3} T_{\beta}(z) G_0(z) U_{\beta 0}(z)
\\
\text{ for } \alpha &= 1, 2, 3,
\end{aligned}
\end{equation}
are Faddeev equations for the three-body transition operators $U_{\alpha \beta}(z)$, where $z$ denotes the three-body energy. The index $\alpha(\beta)$ labels the partitions for the outgoing (incoming) state which consists of three free particles ($\alpha = 0$) or a free particle and dimer ($\alpha = 1, 2, 3$).
$G_0(z)$ is the free resolvent $(z - H_0)^{-1}$, where $H_0$ is the three-body kinetic energy operator for the relative motion. $T_{\alpha}(z)$  represents the two-particle transition operator for the pair $\beta \gamma$ ($\beta,\gamma = 1,2,3$, $\beta \neq \gamma \neq \alpha$), i.e., $T_{\alpha}(z) = V_{\beta \gamma} + V_{\beta \gamma} G_0(z) T_{\alpha}(z)$, where $V_{\beta \gamma}$ is the interaction potential acting within the pair $\beta \gamma$. From the transition amplitude corresponding to $U_{0 0}(0)$, we determine the scattering hypervolume in the same way as described by Ref.~\cite{mestrom2019hypervolumeSqW} which adopts the Weinberg expansion for $T_{\alpha}(z)$ \cite{weinberg1963expansion, mestrom2019squarewell}. 

In the present study, we analyze the scattering hypervolume for identical bosons that interact via various van der Waals potentials, indicated by $V_{\mathrm{LJ}}$, $V_{\mathrm{zero}}$, $V_{\mathrm{exp}}$, and $V_{\mathrm{sc}}$. Their long-range behavior is described by the van der Waals tail $-C_6/r^6$, but their short-range behavior is completely different. Here $V_{\mathrm{LJ}}$ is the Lennard-Jones potential
\begin{equation}
V_{\mathrm{LJ}} = -\frac{C_6}{r^6}\left(1 - \frac{\lambda^6}{r^6} \right),
\end{equation}
where $\lambda$ locates the potential barrier.
The formulas for the other potentials can be found in the Supplemental Material \cite{SupplMat}.  By adjusting the potential depths,
%By adjusting the depth of the potentials,
we tune the scattering length $a$ and the number of two-body bound states \cite{noteBackgroundScatteringLength}. We indicate the number of $s$-wave dimer states by adding an additional index to the potential name. For example, $V_{\mathrm{LJ}}^{(1)}$ supports one $s$-wave dimer state.

\begin{figure}[hbtp]
	\centering
	\includegraphics[width=3.4in]{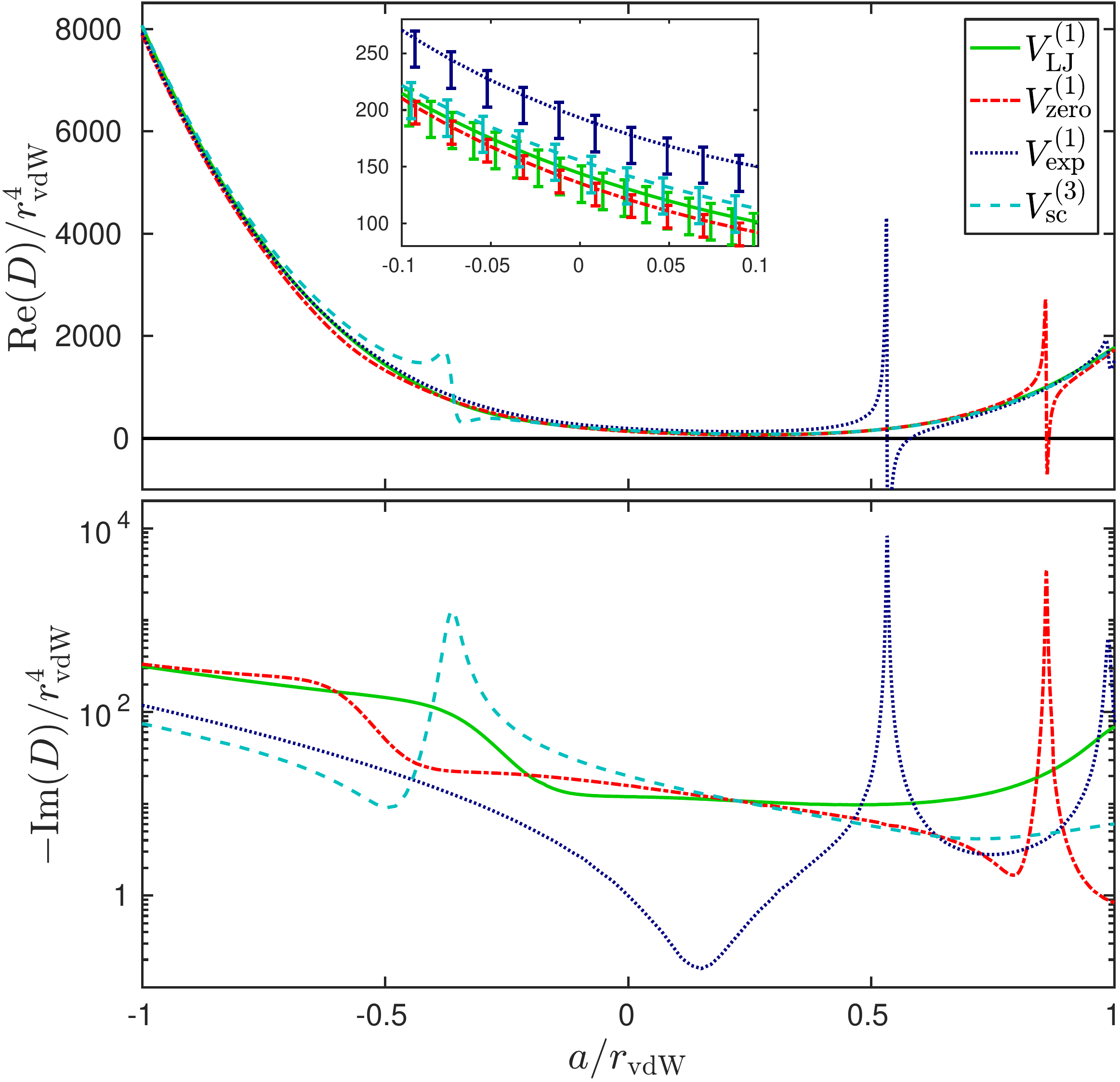}
    \caption{Three-body scattering hypervolume $D$ corresponding to multiple van der Waals potentials as a function of the two-body scattering length $a$ \cite{noteSharpTrimerResonances}. The inset shows the real part of $D$ near the zero crossing of $a$.}
    \label{fig:vdW_Dh_MultiplePotentials}
\end{figure}

\textit{Van der Waals universality.}---Figure \ref{fig:vdW_Dh_MultiplePotentials} shows a comparison of $D$ in the weakly interacting regime for multiple van der Waals potentials. Despite the presence of several three-body resonances, $\mathrm{Re}(D)$ behaves universally in contrast to $\mathrm{Im}(D)$. Physically, $\mathrm{Im}(D)$ is determined by recombination pathways where three atoms approach at short distances, whereas $\mathrm{Re}(D)$ is determined through many competing pathways for elastic scattering at different length scales. When the long-range pathways dominate, $\mathrm{Re}(D)$ is set by the van der Waals tail and the asymptotics of the two-body scattering wave function characterized by $r_{\mathrm{vdW}}$ and $a$, respectively. We note that this picture only applies in the absence of a three-body resonance where the universality in $\mathrm{Re}(D)$ can be broken as shown in \refFigure{fig:vdW_Dh_MultiplePotentials}.

% Figure made using the code: Plot_and_fit_Elastic_K3_List_v34_vdW_comparison_v1.m

%\begin{figure}[hbtp]
%    \centering
%    \includegraphics[width=3.4in]{Figure_Dhyp_Multiple_vdW_a0_-1_1_Dhyp_-1000_8500_ImDhyp_0_4800.pdf}
%    \caption{Three-body scattering hypervolume $D$ corresponding to the multiple van der Waals potentials as a function of the two-body scattering length $a$.  The inset zooms in on the real part of $D$ near the zero-crossing of $a$.}
%    \label{fig:vdW_Dh_MultiplePotentials}
%\end{figure}
% Figure made using the code: Plot_and_fit_Elastic_K3_List_v30_vdW_comparison_v1.m

\begin{figure}[hbtp]
	\centering
	\includegraphics[width=3.4in]{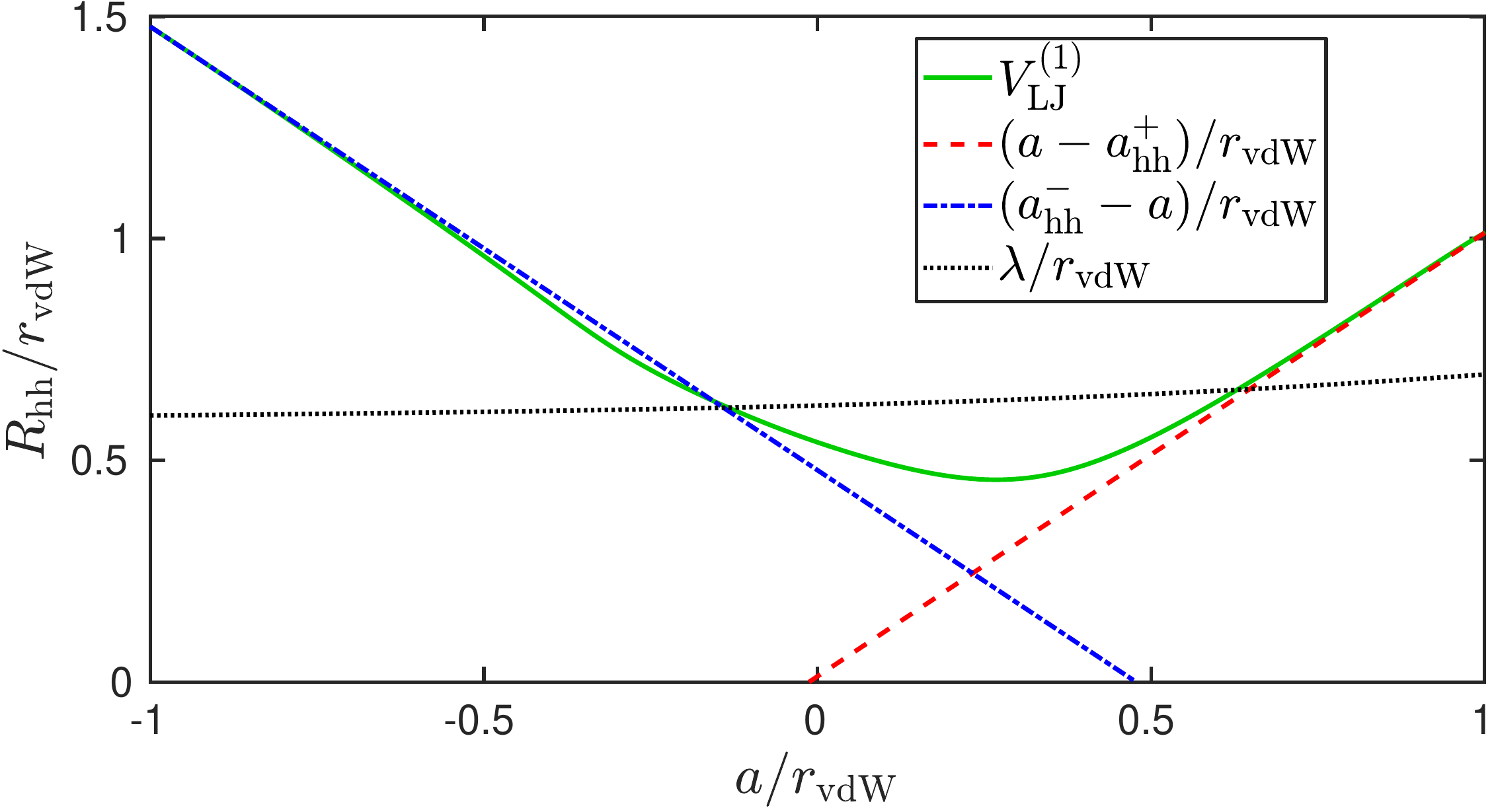}
    \caption{The hard-sphere radius $R_{\mathrm{hh}} = [\mathrm{Re}(D)/1689]^{1/4}$ corresponding to the potential $V_{\mathrm{LJ}}^{(1)}$ (green solid line) as a function of the two-body scattering length $a$. The dashed and dash-dotted curves show $(a - a_{\mathrm{hh}}^{+})/r_{\mathrm{vdW}}$ and $(a_{\mathrm{hh}}^{-} - a)/r_{\mathrm{vdW}}$, respectively, for which $a_{\mathrm{hh}}^{\pm}$ is modified to match $\mathrm{Re}(D)$ at $a/r_{\mathrm{vdW}} = \pm 1$. The values of $a_{\mathrm{hh}}^{+}/r_{\mathrm{vdW}}$ and $a_{\mathrm{hh}}^{-}/r_{\mathrm{vdW}}$ are $-0.012$ and $0.477$, respectively. The dotted curve displays the position $\lambda/r_{\mathrm{vdW}}$ of the repulsive barrier of $V_{\mathrm{LJ}}^{(1)}$. Note that $R_{\mathrm{hh}}$ is slightly affected by a trimer resonance near $a/r_{\mathrm{vdW}}\simeq -0.3$ \cite{SupplMat}.}
    \label{fig:LJvdW_hard_sphere_radius}
\end{figure}

To identify the dominant pathway for elastic three-body scattering, we study the scaling behavior of $\mathrm{Re}(D)$. In the strongly interacting regime ($|a|/r_{\mathrm{vdW}}\gg 1$), pathways that involve a single reflection from the three-body effective potentials contribute to $D$ as $1689 \, a^4$ \cite{dincao2018review, mestrom2019hypervolumeSqW}, both for positive and negative scattering lengths. This reflection occurs off a barrier in the three-body effective potential which acts as a hard hypersphere of hyperradius $|a|$. A similar result $D = 1761.5 \, a^4$ is found for bosons interacting pairwise via a hard-sphere potential \cite{tan2008hypervolume}, whose repulsive character makes it inherently different from the attractive potentials considered in this Letter. However, in the weakly interacting regime ($|a|/r_{\mathrm{vdW}}\lesssim 1$), the scattering length cannot be the only length scale that determines the location of the barrier. Therefore we generalize the hard-hypersphere radius $R_{\mathrm{hh}}$ in the hard-hypersphere formula,
\begin{equation}\label{eq:hard-hypersphere}
\mathrm{Re}(D) = 1689~R_{\mathrm{hh}}^4,
\end{equation}
to $|a~-~a_{\mathrm{hh}}^{\pm}|$, where the offsets $a_{\mathrm{hh}}^{\pm}$ capture finite-range effects. We choose $a_{\mathrm{hh}}^{\pm}$ such that the hard-hypersphere formula matches the value of $\mathrm{Re}(D)$ at $a/r_{\mathrm{vdW}} = \pm 1$. This results in the universal values $a_{\mathrm{hh}}^{+}/r_{\mathrm{vdW}} = -0.010(3)$ and $a_{\mathrm{hh}}^{-}/r_{\mathrm{vdW}} = 0.474(7)$ for which the uncertainties are estimated by the deviation among the considered potentials. That \refEquation{eq:hard-hypersphere} with $R_{\mathrm{hh}} = |a~-~a_{\mathrm{hh}}^{\pm}|$ describes $\mathrm{Re}(D)$ over a range of scattering lengths as shown in Fig.~\ref{fig:LJvdW_hard_sphere_radius} confirms the dominance of hard-hyperspherelike collisions in the weakly interacting regime, except in the crossover regime $-0.1 \lesssim a/r_{\mathrm{vdW}}\lesssim 0.6$. % \del{which we address shortly}.

%Figure~\ref{fig:LJvdW_hard_sphere_radius} shows that the behavior of $\mathrm{Re}(D)$ is also described by the hard-hypersphere formula
%\begin{equation}\label{eq:hard-hypersphere}
%\mathrm{Re}(D) = 1689~R_{\mathrm{hh}}^4,
%\end{equation}
%however now the hard-hypersphere radius $R_{\mathrm{hh}} = |a~-~a_{\mathrm{hh}}^{\pm}|$, where the offsets $a_{\mathrm{hh}}^{\pm}$ are chosen such that the hard-hypersphere formula matches the value of $\mathrm{Re}(D)$ at $a/r_{\mathrm{vdW}} = \pm 1$. That \refEquation{eq:hard-hypersphere} describes $\mathrm{Re}(D)$ over a range of scattering lengths confirms the dominance of hard-hypersphere collisions in the weakly interacting regime, except in the crossover regime $-0.1 \lesssim a/r_{\mathrm{vdW}}\lesssim 0.6$ which we address shortly.

As long as $R_{\mathrm{hh}}>\lambda$, where $\lambda$ is a characteristic length scale of the short-range details of the interaction potential, one can expect that $R_{\mathrm{hh}}$ is set by $a$ and $r_{\mathrm{vdW}}$. Figure~\ref{fig:LJvdW_hard_sphere_radius} shows that the regime where $|a - a_{\mathrm{hh}}^{\pm}|$ does not describe $R_{\mathrm{hh}}$ is roughly determined by $R_{\mathrm{hh}}\lesssim \lambda$. 
%For the Lennard-Jones potential, $\lambda$ locates the potential barrier. 
In the limit of deep van der Waals potentials, $\lambda/r_{\mathrm{vdW}}$ approaches zero. This implies that the $-C_6/r^6$ behavior is approached at smaller values of $r/r_{\mathrm{vdW}}$. Therefore, we expect that $\mathrm{Re}(D)$ of deep van der Waals potentials behaves universally in the complete weakly interacting regime in the absence of trimer resonances \cite{noteResultsSCvdW}.

For a Lennard-Jones potential supporting no two-body bound states ($V_{\mathrm{LJ}}^{(0)}$), we find $D/r_{\mathrm{vdW}}^4 = 86 \pm 2$ at $a~=~0$ and compare this result to the value $90$ predicted by Zwerger \cite{zwerger2019phasetransition,SupplMat,noteZwergerPrediction}.
This prediction is based upon an earlier many-body calculation for the energy per particle of a Bose fluid at zero temperature and zero pressure near the quantum tricritical point  \cite{miller1977phasetransition}. This close agreement demonstrates that $D$ can be determined from properties of ultracold Bose systems near the quantum tricritical point, which we turn to presently.

{\it Thermodynamics and elementary excitations.}---How does the scattering hypervolume determine the ground-state properties and excitations of a BEC near the quantum tricritical point?  To understand these effects, we follow Ref.~\cite{zwerger2019phasetransition} and study the effective Lagrangian density valid  at weak interactions and zero temperature
\begin{align}\label{eq:cubicL}
\mathcal{L}=&\frac{i\hbar}{2}\left[\Psi^*\dot{\Psi}-\Psi\dot{\Psi}^*\right]-\frac{\hbar^2}{2m}|\nabla \Psi|^2-V_\mathrm{ext}({\bf r})|\Psi|^2\nonumber\\
&-\frac{2\pi\hbar^2a}{m}|\Psi|^4-\frac{\hbar^2D}{6m}|\Psi|^6.
\end{align}
The Gross-Pitaevskii equation follows from minimizing the action $S=\int d^3r dt \mathcal{L}$, giving
\begin{equation}\label{eq:cubicgpe}
i\hbar\dot{\Psi}=-\frac{\hbar^2}{2m}\nabla^2\Psi+V_\mathrm{ext}({\bf r})\Psi +\frac{4\pi\hbar^2a}{m}|\Psi|^2\Psi+\frac{\hbar^2D}{2m}|\Psi|^4\Psi.
\end{equation}
The condensate wave function is formulated as $\Psi({\bf r},t)=\sqrt{n({\bf r},t)}e^{-i\phi({\bf r},t)}$ satisfying the Josephson relation $\hbar\dot{\phi}=-\mu$ under stationary conditions. In \refEquation{eq:cubicgpe}, higher-order effects have been ignored as well as the energy
dependence of the few-body scattering amplitudes \cite{gao2003extendedGP,collin2007extendedGP,thogersen2009extendedGP}. %\ins{\refEquation{eq:cubicgpe} could be extended with other higher-order effects like those in the two-body scattering dynamics \cite{gao2003extendedGP,collin2007extendedGP,thogersen2009extendedGP}. However, we neglect such corrections and analyze the stabilizing effect of $D>0$. In particular,} 
In the following, we focus on signatures of $D$ in uniform and trapped systems with $V_\mathrm{ext}({\bf r})= m \omega_{\mathrm{ho}}^2 (\lambda_x^2 x^2+ \lambda_y^2 y^2+ \lambda_z^2 z^2)/2$. We define the oscillator length $l_{\mathrm{ho}} = \sqrt{\hbar/m\omega_\mathrm{ho}}$ and the geometric means $\bar{\omega} = \omega_{\mathrm{ho}} (\lambda_x \lambda_y \lambda_z)^{1/3}$ and $\bar{l} = \sqrt{\hbar/m\bar{\omega}}$. In this analysis, we neglect the imaginary part of $D$ since our results for $V_{\mathrm{LJ}}^{(1)}$ show that $|\mathrm{Re}(D)/\mathrm{Im}(D)|$ varies from roughly 10 near $a = 0$ to roughly 25 near $a/r_{\mathrm{vdW}} = \pm 1$, and we return to this point below when discussing the experimental outlook. 

At the point $a=0$, where we estimate $\mathrm{Re}(D)/r_{\mathrm{vdW}}^4 \approx 100$, Eqs.~\eqref{eq:cubicL} and \eqref{eq:cubicgpe} contain only effective three-body interactions. This regime for a uniform gas was considered in Ref.~\cite{zwerger2019phasetransition}, finding chemical potential $\mu=D\hbar^2n^2/2m$, pressure $P=(8m/9D\hbar^2)^{1/2}\mu^{3/2}$ and sound velocity $c = \hbar n \sqrt{D}/m$.
%finding chemical potential $\mu=D\hbar^2n^2/2m$ and \ins{pressure} $P=(8m/9D\hbar^2)^{1/2}\mu^{3/2}$.  Sound waves propagate with velocity $c$ that is determined by \del{$\partial_n P=mc^2=\hbar^2Dn^2/m$} \cite{zwerger2019phasetransition}.
For the ground state of a trapped gas, the Thomas-Fermi approximation gives $n({\bf r})=(2m[\mu-V_\mathrm{ext}({\bf r})]/D\hbar^2)^{1/2}$.
% the local density approximation gives the Thomas-Fermi profile $n({\bf r})=(2m(\mu-V_\mathrm{ext}({\bf r}))/D\hbar^2)^{1/2}$.   
After normalization, the chemical potential is fixed to $\mu/\hbar \bar{\omega} =\zeta^{1/4}/\pi$ in terms of the three-body Thomas-Fermi parameter $\zeta= DN^2/\bar{l}^4$.  Likewise, integrating the thermodynamic relation $\mu=\partial_N E$ gives energy per particle $E/N=2\mu/3$, from which we infer the interaction energy per particle $E_\mathrm{int}/N=\mu/6$ as a consequence of the virial theorem \cite{RevModPhys.71.463}.  At the cloud boundaries $\mu=m\omega_{\mathrm{ho}}^2 \lambda_{\eta}^2 R_{\eta}^2/2$ (${\eta} = x, y, z$), and we estimate the spatial extent of the cloud from the geometric mean of the semi-axes $\bar{R}\equiv(R_xR_yR_z)^{1/3} = \sqrt{2/\pi} \ \bar{l}\zeta^{1/8}$.  Compared to the Thomas-Fermi limit for two-body interactions, we find that a smaller portion of the total energy is involved in interactions, however, due to the $N^2$ scaling of $\zeta$, all energies and radii scale with higher powers of $N$.

\begin{figure}[t!]
	\centering
	\includegraphics[width=3.4in]{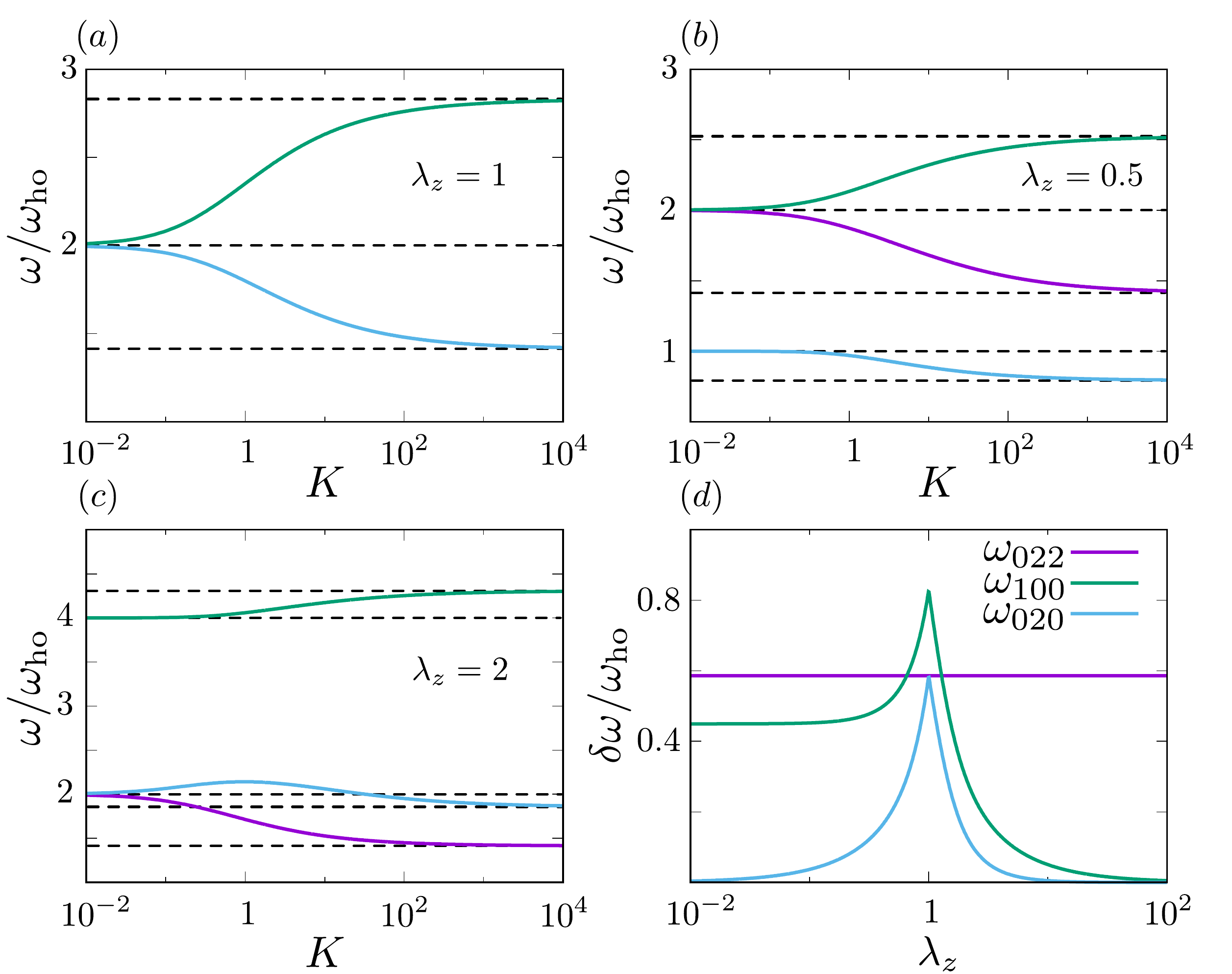}
    \caption{Collective mode frequencies at $a=0$ versus $K~=~2 D N^2/9\sqrt{3}\pi^3 l_{\mathrm{ho}}^4$ for (a) isotropic, (b) cigar, and (c) pancake geometries. We set $\lambda_x = \lambda_y = 1$ and vary $\lambda_z$. The $\omega_{020}$ and $\omega_{022}$ modes are degenerate for an isotropic trap.  Dashed lines indicate results in the noninteracting and Thomas-Fermi limits.  (d) Contrast $\delta\omega_{nlm}\equiv\omega_{nlm}|_{K\gg1}-\omega_{nlm}|_{K=0}$ versus trap aspect ratio $\lambda_z$.}
    \label{fig:trapmodes}
\end{figure}

To investigate the shift of discretized collective modes  in a harmonic trap at $a=0$, we use a time-dependent trial wave function \cite{PhysRevLett.77.5320,PhysRevA.63.053607,Al_Jibbouri_2013}
\begin{equation}\label{eq:Gaussian_ansatz_excitations}
\Psi(x,y,z,t)= A(t) \prod_{\eta=x,y,z}e^{-\frac{(\eta-\eta_0)^2}{2w_\eta^2}+i\eta\alpha_\eta+i\eta^2\beta_\eta},
\end{equation}
with time-dependent variational parameters $\{w_\eta, \eta_0,\alpha_\eta,\beta_\eta\}_{\eta=x,y,z}$.  The magnitude of $A$ is set by particle number conservation. Minimizing the action with respect to the variational parameters, we find 
\begin{align}
&\partial_\tau^2\eta_0+\lambda_\eta^2\eta_0=0,\label{eq:com}\\
&\partial^2_\tau  v_{\eta}+\lambda_{\eta}^2 v_{\eta}=\frac{1}{v_{\eta}^3}+\frac{K}{v_{\eta} (v_x v_y v_z)^2} \quad (\eta = x, y, z),\label{eq:widths}
\end{align}
with $K=2 D N^2/9\sqrt{3}\pi^3 l_{\mathrm{ho}}^4$ and dimensionless scalings $v_\eta=w_\eta/l_\mathrm{ho}$ and $\tau=\omega_\mathrm{ho} t$.  Equation~\eqref{eq:com} describes dipole oscillations of the condensate center with trap frequencies in agreement with the Kohn theorem \cite{PhysRev.123.1242}.  Linearizing Eq.~\eqref{eq:widths} about equilibrium, yields mode frequencies $\omega_{022}$, $\omega_{100}$, and $\omega_{020}$  in terms of principle and angular quantum numbers parametrized as $\omega_{nlm}$ \cite{PhysRevLett.77.5320} (see Ref.~\cite{SupplMat}).  Our results for these modes at $a=0$ are shown in Figs.~\ref{fig:trapmodes}(a)-\ref{fig:trapmodes}(d) over a range of geometries and values of $K$ compared against results in the noninteracting and Thomas-Fermi limits.  Experimentally, $D$ can be inferred from frequency shifts intermediate to these limits.  From Fig.~\ref{fig:trapmodes}(d), this contrast is maximized for the breathing mode $\omega_{100}$ in an isotropic geometry. In the Thomas-Fermi limit, this mode shifts to  $2\sqrt{2}\omega_\mathrm{ho}$, which is beyond the result $\sqrt{5}\omega_\mathrm{ho}$ for two-body interactions \cite{PhysRevLett.77.2360}. Physically, the simultaneous compression of this mode along all axes leads to increased densities near trap center and the largest nonlinear interaction effects \cite{PhysRevA.94.043640}.  

\begin{figure}[t!]
	\centering
	\includegraphics[width=3.4in]{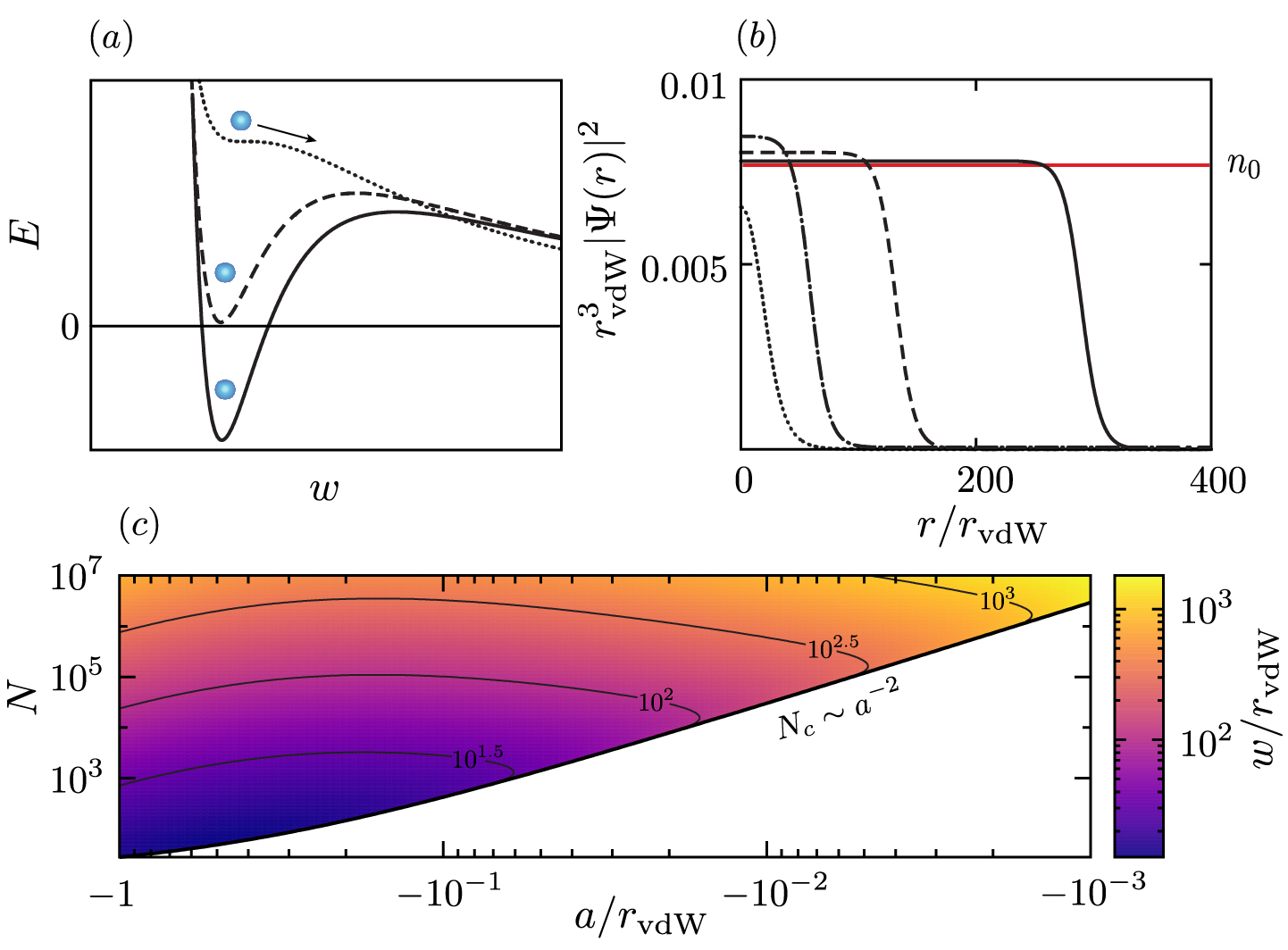}
    \caption{Ground-state properties of Bose droplets.  (a) Illustrated dependence of the energy $E$ given by \refEquation{eq:Energy_Gauss_No_trap} with width $w$ for $N> N_{\mathrm{s}}$ (solid), $N = N_{\mathrm{s}}$ (dashed), and $N = N_{\mathrm{c}}$ (dotted).  (b)  Numerical ground-state profile versus radial coordinate for $N/N_{\mathrm{s}}\approx 1$ (dotted), $10$ (dash-dotted),  $100$ (dashed), and $1000$ (solid) evaluated at $a=-0.08~r_\mathrm{vdW}$ taking our result for $V_{\mathrm{LJ}}^{(1)}$, $\mathrm{Re}(D)=197~r_\mathrm{vdW}^4$ (see \refFigure{fig:vdW_Dh_MultiplePotentials} inset), compared with the large-$N$ density $n_0$ indicated by the solid red curve. (c) Droplet width $w$ versus $a$ and $N$ from the Gaussian trial wave function Eq.~\eqref{eq:Gaussian_ansatz_excitations} using \refEquation{eq:hard-hypersphere} with $R_{\mathrm{hh}} = |a - 0.477~r_{\mathrm{vdW}}|$ and neglecting $\mathrm{Im}(D)$. The liquid-to-gas transition at $N_{\mathrm{c}}$ is indicated by the solid line.}
    \label{fig:droplets}
\end{figure}

For $a<0$, it is possible that $D>0$ can stabilize the BEC against collapse. Taking the Gaussian trial wave function of \refEquation{eq:Gaussian_ansatz_excitations} with $\eta_0 = \alpha_\eta = \beta_{\eta} = 0$ and $w_{\eta} = w$,
%\begin{equation}\label{eq:Gaussian_ansatz}
%\psi(\mathbf{r}) = \sqrt{\frac{N}{\pi^{3/2} w^3}} e^{-\frac{r^2}{2 w^2}},
%\end{equation}
we minimize the energy 
\begin{equation} \label{eq:Energy_Gauss_No_trap}
\begin{aligned}
E &= \frac{3 \hbar^2}{4 m} \frac{N}{w^2} 
+
\frac{\hbar^2 a}{\sqrt{2 \pi} m} 
\frac{N^2}{w^3}
+
\frac{\hbar^2 D}{18 \sqrt{3} \pi^{3} m} \frac{N^3}{w^6}
\end{aligned}
\end{equation}
in the absence of a trapping potential for a fixed number of particles $N$. We find that no droplets exist for $N \leq N_{\mathrm{c}}$ where
\begin{equation}\label{eq:Nc_Gauss_metastable}
\begin{aligned}
N_{\mathrm{c}} &= \frac{2^{7/2}}{3^{11/4} \sqrt{\pi}} \frac{\sqrt{D}}{a^2}.
\end{aligned}
\end{equation}
The droplets are metastable for $N_{\mathrm{c}} < N \leq 3 \sqrt{3} N_{\mathrm{c}}/4  = N_{\mathrm{s}}$ and stable for $N > N_{\mathrm{s}}$. The variational dependence of the energy as a function of the width $w$ is illustrated in \refFigure{fig:droplets}(a) for the unstable, metastable, and stable regimes. The density profiles of the droplets numerically obtained from \refEquation{eq:cubicgpe} are depicted in \refFigure{fig:droplets}(b) for various $N$ which shows that the Gaussian trial wave function is reasonable near the metastable regime. Figure~ \ref{fig:droplets}(c) shows the phase diagram of the Bose fluid for $a<0$ using \refEquation{eq:hard-hypersphere} with $R_{\mathrm{hh}} = |a - 0.477~r_{\mathrm{vdW}}|$ while neglecting $\mathrm{Im}(D)$.
For large $N$ the density profile is almost constant with density $n_0 = 6 \pi |a|/D$ \cite{bulgac2002droplets} which is approximately a factor $2.18$ larger than the center density of a droplet with $N = N_{\mathrm{c}}$. In the large $N$ limit, the Gaussian trial wave function overestimates the center density by roughly a factor 1.84.
Previous estimates of $n_0$ in Ref.~\cite{gao2004dropletsvdW} 
% for $a/r_{\mathrm{vdW}} = -1.22\cdot 10^{-2}$ to $-0.366$ 
are larger by roughly a factor of 5 to 10 than our result, which we attribute to an underestimation of effective three-body repulsion in that work. 

\textit{Experimental outlook.}---Let us now discuss the experimental possibilities to observe the effects of $D$. For this purpose, we take $\text{Im}(D)$ into account in the typical lifetime $\tau_{\mathrm{life}} = 1/L_3 n^2$ where $L_3 = -\text{Im}(D) \hbar/m$ determines the loss rate of atoms from the BEC via three-body recombination \cite{tan2017hypervolume}. For quantum droplets with density $n_0$ and chemical potential $\mu_0 = -6 \pi^2 \hbar^2 a^2/m \text{Re}(D)$ \cite{bulgac2002droplets}, we compare $\tau_{\mathrm{life}}$ to the characteristic timescale $\tau_0 = \hbar/|\mu_0|$ and find $\tau_{\mathrm{life}}/\tau_{0} = -\mathrm{Re}(D)/6\mathrm{Im}(D)$,
%\begin{equation}
%\frac{\tau_{\mathrm{life}}}{\tau_{0}} = -\frac{1}{6} \frac{\mathrm{Re}(D)}{\mathrm{Im}(D)},
%\end{equation}
which is typically larger than one according to our results for $D$. Since $\tau_{\mathrm{life}} \propto [\mathrm{Re}(D)]^2/\mathrm{Im}(D) a^2$, longer lifetimes can be achieved at smaller $|a|$. However, the critical number $N_{\mathrm{c}}$ also increases in this limit [see \refFigure{fig:droplets}(c)].

%For example, to obtain a droplet with $\tau_{\text{life}} = 100$~ms we estimate that $a$ should be on the order of  $-10^{-3}~r_{\mathrm{vdW}}$ and $N$ on the order of $10^7$ (see \refFigure{fig:droplets}(c)).
%\begin{equation}
%\tau_{\mathrm{life}} = -\frac{1}{36 \pi^2} \frac{m}{\hbar} \frac{\mathrm{Re}^2(D)}{\mathrm{Im}(D) a^2}
%\end{equation}

%since the droplet is unstable for $N$ below the critical atom number $N_{\mathrm{c}}$ which we again determine variationally for a Gaussian density profile leading to 
%\begin{equation*}
%\begin{aligned}
%N_{\mathrm{c}} &= \frac{8 \sqrt{\frac{2}{\pi}}}{9\cdot 3^{3/4}} \frac{\sqrt{\mathrm{Re}(D)}}{a^2}
%\\
%&\simeq 0.3111 \frac{\sqrt{\mathrm{Re}(D)}}{a^2}.
%\end{aligned}
%\end{equation*}

To observe the collective mode frequencies at $a = 0$, we require $\tau_{\mathrm{life}} > \tau_{\mathrm{ho}} = 2 \pi/\omega_{\mathrm{ho}}$. In the Thomas-Fermi limit, we find
\begin{equation} \label{eq:taulife_tauho_vs_K_lambda}
\begin{aligned}
\frac{\tau_{\mathrm{life}}}{\tau_{\mathrm{ho}}}
&= -\frac{\mathrm{Re}(D)}{\mathrm{Im}(D)}\frac{1}{2 \cdot 3^{5/8} (2 \pi)^{3/4} \lambda_z^{1/2} K^{1/4}}
\end{aligned}
\end{equation}
for cylindrically symmetric traps with $\lambda_x = \lambda_y = 1$ using the center density to estimate $\tau_{\mathrm{life}}$. In general, smaller $\lambda_z$ is advantageous for achieving large $K$ for fixed $\tau_{\mathrm{life}}/\tau_{\mathrm{ho}}$ and $D$. Figure~\ref{fig:trapmodes}(d) shows that $\delta \omega_{022}/\omega_{\mathrm{ho}}$ and $\delta \omega_{100}/\omega_{\mathrm{ho}}$ are roughly 0.5 for small $\lambda_z$ which makes these modes suitable for extracting $\mathrm{Re}(D)$ from the $K$ dependence of the corresponding mode frequencies. Cigar-shaped traps can also be used to measure the speed of sound at $a = 0$ provided that the characteristic distance $c \tau_{\mathrm{life}}$ is large enough to resolve experimentally. Using our previous estimate for $\tau_{\mathrm{life}}$ in the Thomas-Fermi limit, we find 
\begin{equation}
c \tau_{\mathrm{life}} = -\sqrt{\frac{\pi}{2}} \frac{[\mathrm{Re}(D)]^{5/4}}{\mathrm{Im}(D)} \frac{\sqrt{N}}{\zeta^{3/8}},
\end{equation}
which scales as $\lambda_z^{-1/4}$.

\textit{Conclusion.}---We study the scattering hypervolume $D$ for identical bosons interacting via pairwise van der Waals potentials. Our results show that $\mathrm{Re}(D)$ is predominantly determined by the long-range two-body properties $r_{\mathrm{vdW}}$ and $a$. The van der Waals universality of this behavior is due to dominant hard-hypersphere scattering. However, $\mathrm{Im}(D)$ depends strongly on the short-range details of the interaction, resulting in nonuniversal behavior. In the limit of vanishing $a$, $\mathrm{Re}(D)$ determines the quantum phase diagram near the tricritical point. Using the van der Waals universality of $\mathrm{Re}(D)$, we make quantitative predictions for the properties of atomic BECs, including the formation of quantum droplets.

Further studies of $D$ for deeper van der Waals potentials need to be conducted to test the robustness of the universal behavior in $\mathrm{Re}(D)$ in the weakly interacting regime. The influence of multichannel physics on $D$ could lead to new interesting phenomena at small scattering lengths due to additional parameters characterizing the zero crossing of $a$ \cite{khaykovich2014zeroa0, petrov2019zeroScatL3bodyInteraction,chin2010feshbach}.
Dynamical studies including effective three-body repulsion are needed to understand droplet formation.

\textit{Acknowledgements.}---We thank Wilhelm Zwerger, Jinglun Li, Silvia Musolino, and Denise Ahmed-Braun for fruitful discussions. This research is financially supported by the Netherlands Organisation for Scientific Research (NWO) under Grant No. 680-47-623 and by the Foundation for Fundamental Research on Matter (FOM).

\bibliographystyle{apsrev}
\bibliography{Bibliography}

\begin{thebibliography}{57}
\expandafter\ifx\csname natexlab\endcsname\relax\def\natexlab#1{#1}\fi
\expandafter\ifx\csname bibnamefont\endcsname\relax
  \def\bibnamefont#1{#1}\fi
\expandafter\ifx\csname bibfnamefont\endcsname\relax
  \def\bibfnamefont#1{#1}\fi
\expandafter\ifx\csname citenamefont\endcsname\relax
  \def\citenamefont#1{#1}\fi
\expandafter\ifx\csname url\endcsname\relax
  \def\url#1{\texttt{#1}}\fi
\expandafter\ifx\csname urlprefix\endcsname\relax\def\urlprefix{URL }\fi
\providecommand{\bibinfo}[2]{#2}
\providecommand{\eprint}[2][]{\url{#2}}

\bibitem[{\citenamefont{Huang}(1963)}]{huang1963statistical}
\bibinfo{author}{\bibfnamefont{K.}~\bibnamefont{Huang}},
  \emph{\bibinfo{title}{Statistical Mechanics}} (\bibinfo{publisher}{Wiley},
  \bibinfo{address}{New York}, \bibinfo{year}{1963}).

\bibitem[{\citenamefont{Tan}(2008)}]{tan2008hypervolume}
\bibinfo{author}{\bibfnamefont{S.}~\bibnamefont{Tan}}, \bibinfo{journal}{Phys.
  Rev. A} \textbf{\bibinfo{volume}{78}}, \bibinfo{pages}{013636}
  (\bibinfo{year}{2008}).

\bibitem[{\citenamefont{Zwerger}(2019)}]{zwerger2019phasetransition}
\bibinfo{author}{\bibfnamefont{W.}~\bibnamefont{Zwerger}},
  \bibinfo{journal}{Journal of Statistical Mechanics: Theory and Experiment}
  \textbf{\bibinfo{volume}{2019}}, \bibinfo{pages}{103104}
  (\bibinfo{year}{2019}).

\bibitem[{\citenamefont{Belitz and Kirkpatrick}(2017)}]{belitz2017tricritical}
\bibinfo{author}{\bibfnamefont{D.}~\bibnamefont{Belitz}} \bibnamefont{and}
  \bibinfo{author}{\bibfnamefont{T.~R.} \bibnamefont{Kirkpatrick}},
  \bibinfo{journal}{Phys. Rev. Lett.} \textbf{\bibinfo{volume}{119}},
  \bibinfo{pages}{267202} (\bibinfo{year}{2017}).

\bibitem[{\citenamefont{Friedemann et~al.}(2018)\citenamefont{Friedemann,
  Duncan, Hirschberger, Bauer, K{\"u}chler, Neubauer, Brando, Pfleiderer, and
  Grosche}}]{friedemann2018tricritical}
\bibinfo{author}{\bibfnamefont{S.}~\bibnamefont{Friedemann}},
  \bibinfo{author}{\bibfnamefont{W.~J.} \bibnamefont{Duncan}},
  \bibinfo{author}{\bibfnamefont{M.}~\bibnamefont{Hirschberger}},
  \bibinfo{author}{\bibfnamefont{T.~W.} \bibnamefont{Bauer}},
  \bibinfo{author}{\bibfnamefont{R.}~\bibnamefont{K{\"u}chler}},
  \bibinfo{author}{\bibfnamefont{A.}~\bibnamefont{Neubauer}},
  \bibinfo{author}{\bibfnamefont{M.}~\bibnamefont{Brando}},
  \bibinfo{author}{\bibfnamefont{C.}~\bibnamefont{Pfleiderer}},
  \bibnamefont{and} \bibinfo{author}{\bibfnamefont{F.~M.}
  \bibnamefont{Grosche}}, \bibinfo{journal}{Nature Physics}
  \textbf{\bibinfo{volume}{14}}, \bibinfo{pages}{62} (\bibinfo{year}{2018}).

\bibitem[{\citenamefont{Kaluarachchi et~al.}(2018)\citenamefont{Kaluarachchi,
  Taufour, Bud'ko, and Canfield}}]{canfield2018tricritical}
\bibinfo{author}{\bibfnamefont{U.~S.} \bibnamefont{Kaluarachchi}},
  \bibinfo{author}{\bibfnamefont{V.}~\bibnamefont{Taufour}},
  \bibinfo{author}{\bibfnamefont{S.~L.} \bibnamefont{Bud'ko}},
  \bibnamefont{and} \bibinfo{author}{\bibfnamefont{P.~C.}
  \bibnamefont{Canfield}}, \bibinfo{journal}{Phys. Rev. B}
  \textbf{\bibinfo{volume}{97}}, \bibinfo{pages}{045139}
  (\bibinfo{year}{2018}).

\bibitem[{\citenamefont{Xu and Pu}(2019)}]{xu2019tricritical}
\bibinfo{author}{\bibfnamefont{Y.}~\bibnamefont{Xu}} \bibnamefont{and}
  \bibinfo{author}{\bibfnamefont{H.}~\bibnamefont{Pu}}, \bibinfo{journal}{Phys.
  Rev. Lett.} \textbf{\bibinfo{volume}{122}}, \bibinfo{pages}{193201}
  (\bibinfo{year}{2019}).

\bibitem[{\citenamefont{Bulgac}(2002)}]{bulgac2002droplets}
\bibinfo{author}{\bibfnamefont{A.}~\bibnamefont{Bulgac}},
  \bibinfo{journal}{Phys. Rev. Lett.} \textbf{\bibinfo{volume}{89}},
  \bibinfo{pages}{050402} (\bibinfo{year}{2002}).

\bibitem[{\citenamefont{Gao}(2004)}]{gao2004dropletsvdW}
\bibinfo{author}{\bibfnamefont{B.}~\bibnamefont{Gao}},
  \bibinfo{journal}{Journal of Physics B: Atomic, Molecular and Optical
  Physics} \textbf{\bibinfo{volume}{37}}, \bibinfo{pages}{L227}
  (\bibinfo{year}{2004}).

\bibitem[{\citenamefont{Gerton et~al.}(2000)\citenamefont{Gerton, Strekalov,
  Prodan, and Hulet}}]{gerton2000collapseBEC}
\bibinfo{author}{\bibfnamefont{J.~M.} \bibnamefont{Gerton}},
  \bibinfo{author}{\bibfnamefont{D.}~\bibnamefont{Strekalov}},
  \bibinfo{author}{\bibfnamefont{I.}~\bibnamefont{Prodan}}, \bibnamefont{and}
  \bibinfo{author}{\bibfnamefont{R.~G.} \bibnamefont{Hulet}},
  \bibinfo{journal}{Nature} \textbf{\bibinfo{volume}{408}},
  \bibinfo{pages}{692} (\bibinfo{year}{2000}).

\bibitem[{\citenamefont{Roberts et~al.}(2001)\citenamefont{Roberts, Claussen,
  Cornish, Donley, Cornell, and Wieman}}]{roberts2001collapseBEC}
\bibinfo{author}{\bibfnamefont{J.~L.} \bibnamefont{Roberts}},
  \bibinfo{author}{\bibfnamefont{N.~R.} \bibnamefont{Claussen}},
  \bibinfo{author}{\bibfnamefont{S.~L.} \bibnamefont{Cornish}},
  \bibinfo{author}{\bibfnamefont{E.~A.} \bibnamefont{Donley}},
  \bibinfo{author}{\bibfnamefont{E.~A.} \bibnamefont{Cornell}},
  \bibnamefont{and} \bibinfo{author}{\bibfnamefont{C.~E.}
  \bibnamefont{Wieman}}, \bibinfo{journal}{Phys. Rev. Lett.}
  \textbf{\bibinfo{volume}{86}}, \bibinfo{pages}{4211} (\bibinfo{year}{2001}).

\bibitem[{\citenamefont{Donley et~al.}(2001)\citenamefont{Donley, Claussen,
  Cornish, Roberts, Cornell, and Wieman}}]{donley2001collapseBEC}
\bibinfo{author}{\bibfnamefont{E.~A.} \bibnamefont{Donley}},
  \bibinfo{author}{\bibfnamefont{N.~R.} \bibnamefont{Claussen}},
  \bibinfo{author}{\bibfnamefont{S.~L.} \bibnamefont{Cornish}},
  \bibinfo{author}{\bibfnamefont{J.~L.} \bibnamefont{Roberts}},
  \bibinfo{author}{\bibfnamefont{E.~A.} \bibnamefont{Cornell}},
  \bibnamefont{and} \bibinfo{author}{\bibfnamefont{C.~E.}
  \bibnamefont{Wieman}}, \bibinfo{journal}{Nature}
  \textbf{\bibinfo{volume}{412}}, \bibinfo{pages}{295} (\bibinfo{year}{2001}).

\bibitem[{\citenamefont{Eigen et~al.}(2016)\citenamefont{Eigen, Gaunt,
  Suleymanzade, Navon, Hadzibabic, and Smith}}]{eigen2016collapseBEC}
\bibinfo{author}{\bibfnamefont{C.}~\bibnamefont{Eigen}},
  \bibinfo{author}{\bibfnamefont{A.~L.} \bibnamefont{Gaunt}},
  \bibinfo{author}{\bibfnamefont{A.}~\bibnamefont{Suleymanzade}},
  \bibinfo{author}{\bibfnamefont{N.}~\bibnamefont{Navon}},
  \bibinfo{author}{\bibfnamefont{Z.}~\bibnamefont{Hadzibabic}},
  \bibnamefont{and} \bibinfo{author}{\bibfnamefont{R.~P.} \bibnamefont{Smith}},
  \bibinfo{journal}{Phys. Rev. X} \textbf{\bibinfo{volume}{6}},
  \bibinfo{pages}{041058} (\bibinfo{year}{2016}).

\bibitem[{\citenamefont{Petrov}(2015)}]{petrov2015dropletsBoseBoseMix}
\bibinfo{author}{\bibfnamefont{D.~S.} \bibnamefont{Petrov}},
  \bibinfo{journal}{Phys. Rev. Lett.} \textbf{\bibinfo{volume}{115}},
  \bibinfo{pages}{155302} (\bibinfo{year}{2015}).

\bibitem[{\citenamefont{Cabrera et~al.}(2018)\citenamefont{Cabrera, Tanzi,
  Sanz, Naylor, Thomas, Cheiney, and Tarruell}}]{cabrera2018quantumdroplet}
\bibinfo{author}{\bibfnamefont{C.~R.} \bibnamefont{Cabrera}},
  \bibinfo{author}{\bibfnamefont{L.}~\bibnamefont{Tanzi}},
  \bibinfo{author}{\bibfnamefont{J.}~\bibnamefont{Sanz}},
  \bibinfo{author}{\bibfnamefont{B.}~\bibnamefont{Naylor}},
  \bibinfo{author}{\bibfnamefont{P.}~\bibnamefont{Thomas}},
  \bibinfo{author}{\bibfnamefont{P.}~\bibnamefont{Cheiney}}, \bibnamefont{and}
  \bibinfo{author}{\bibfnamefont{L.}~\bibnamefont{Tarruell}},
  \bibinfo{journal}{Science} \textbf{\bibinfo{volume}{359}},
  \bibinfo{pages}{301} (\bibinfo{year}{2018}).

\bibitem[{\citenamefont{Cheiney et~al.}(2018)\citenamefont{Cheiney, Cabrera,
  Sanz, Naylor, Tanzi, and Tarruell}}]{tarruell2018quantumdroplet}
\bibinfo{author}{\bibfnamefont{P.}~\bibnamefont{Cheiney}},
  \bibinfo{author}{\bibfnamefont{C.~R.} \bibnamefont{Cabrera}},
  \bibinfo{author}{\bibfnamefont{J.}~\bibnamefont{Sanz}},
  \bibinfo{author}{\bibfnamefont{B.}~\bibnamefont{Naylor}},
  \bibinfo{author}{\bibfnamefont{L.}~\bibnamefont{Tanzi}}, \bibnamefont{and}
  \bibinfo{author}{\bibfnamefont{L.}~\bibnamefont{Tarruell}},
  \bibinfo{journal}{Phys. Rev. Lett.} \textbf{\bibinfo{volume}{120}},
  \bibinfo{pages}{135301} (\bibinfo{year}{2018}).

\bibitem[{\citenamefont{Semeghini et~al.}(2018)\citenamefont{Semeghini,
  Ferioli, Masi, Mazzinghi, Wolswijk, Minardi, Modugno, Modugno, Inguscio, and
  Fattori}}]{semeghini2018quantumdroplet}
\bibinfo{author}{\bibfnamefont{G.}~\bibnamefont{Semeghini}},
  \bibinfo{author}{\bibfnamefont{G.}~\bibnamefont{Ferioli}},
  \bibinfo{author}{\bibfnamefont{L.}~\bibnamefont{Masi}},
  \bibinfo{author}{\bibfnamefont{C.}~\bibnamefont{Mazzinghi}},
  \bibinfo{author}{\bibfnamefont{L.}~\bibnamefont{Wolswijk}},
  \bibinfo{author}{\bibfnamefont{F.}~\bibnamefont{Minardi}},
  \bibinfo{author}{\bibfnamefont{M.}~\bibnamefont{Modugno}},
  \bibinfo{author}{\bibfnamefont{G.}~\bibnamefont{Modugno}},
  \bibinfo{author}{\bibfnamefont{M.}~\bibnamefont{Inguscio}}, \bibnamefont{and}
  \bibinfo{author}{\bibfnamefont{M.}~\bibnamefont{Fattori}},
  \bibinfo{journal}{Phys. Rev. Lett.} \textbf{\bibinfo{volume}{120}},
  \bibinfo{pages}{235301} (\bibinfo{year}{2018}).

\bibitem[{\citenamefont{Kadau et~al.}(2016)\citenamefont{Kadau, Schmitt,
  Wenzel, Wink, Maier, Ferrier-Barbut, and
  Pfau}}]{kadau2016dipolarquantumdroplet}
\bibinfo{author}{\bibfnamefont{H.}~\bibnamefont{Kadau}},
  \bibinfo{author}{\bibfnamefont{M.}~\bibnamefont{Schmitt}},
  \bibinfo{author}{\bibfnamefont{M.}~\bibnamefont{Wenzel}},
  \bibinfo{author}{\bibfnamefont{C.}~\bibnamefont{Wink}},
  \bibinfo{author}{\bibfnamefont{T.}~\bibnamefont{Maier}},
  \bibinfo{author}{\bibfnamefont{I.}~\bibnamefont{Ferrier-Barbut}},
  \bibnamefont{and} \bibinfo{author}{\bibfnamefont{T.}~\bibnamefont{Pfau}},
  \bibinfo{journal}{Nature} \textbf{\bibinfo{volume}{530}},
  \bibinfo{pages}{194} (\bibinfo{year}{2016}).

\bibitem[{\citenamefont{Ferrier-Barbut
  et~al.}(2016)\citenamefont{Ferrier-Barbut, Kadau, Schmitt, Wenzel, and
  Pfau}}]{ferrierbarbut2016dipolarquantumdroplet}
\bibinfo{author}{\bibfnamefont{I.}~\bibnamefont{Ferrier-Barbut}},
  \bibinfo{author}{\bibfnamefont{H.}~\bibnamefont{Kadau}},
  \bibinfo{author}{\bibfnamefont{M.}~\bibnamefont{Schmitt}},
  \bibinfo{author}{\bibfnamefont{M.}~\bibnamefont{Wenzel}}, \bibnamefont{and}
  \bibinfo{author}{\bibfnamefont{T.}~\bibnamefont{Pfau}},
  \bibinfo{journal}{Phys. Rev. Lett.} \textbf{\bibinfo{volume}{116}},
  \bibinfo{pages}{215301} (\bibinfo{year}{2016}).

\bibitem[{\citenamefont{Chomaz et~al.}(2016)\citenamefont{Chomaz, Baier,
  Petter, Mark, W\"achtler, Santos, and
  Ferlaino}}]{chomaz2016dipolarquantumdroplet}
\bibinfo{author}{\bibfnamefont{L.}~\bibnamefont{Chomaz}},
  \bibinfo{author}{\bibfnamefont{S.}~\bibnamefont{Baier}},
  \bibinfo{author}{\bibfnamefont{D.}~\bibnamefont{Petter}},
  \bibinfo{author}{\bibfnamefont{M.~J.} \bibnamefont{Mark}},
  \bibinfo{author}{\bibfnamefont{F.}~\bibnamefont{W\"achtler}},
  \bibinfo{author}{\bibfnamefont{L.}~\bibnamefont{Santos}}, \bibnamefont{and}
  \bibinfo{author}{\bibfnamefont{F.}~\bibnamefont{Ferlaino}},
  \bibinfo{journal}{Phys. Rev. X} \textbf{\bibinfo{volume}{6}},
  \bibinfo{pages}{041039} (\bibinfo{year}{2016}).

\bibitem[{\citenamefont{Zhu and Tan}(2017)}]{tan2017hypervolume}
\bibinfo{author}{\bibfnamefont{S.}~\bibnamefont{Zhu}} \bibnamefont{and}
  \bibinfo{author}{\bibfnamefont{S.}~\bibnamefont{Tan}},
  \bibinfo{journal}{arXiv:1710.04147v1 [cond-mat.quant-gas]}
  (\bibinfo{year}{2017}).

\bibitem[{\citenamefont{Braaten and Hammer}(2006)}]{braaten2006universality}
\bibinfo{author}{\bibfnamefont{E.}~\bibnamefont{Braaten}} \bibnamefont{and}
  \bibinfo{author}{\bibfnamefont{H.-W.} \bibnamefont{Hammer}},
  \bibinfo{journal}{Phys. Rep.} \textbf{\bibinfo{volume}{428}},
  \bibinfo{pages}{259} (\bibinfo{year}{2006}).

\bibitem[{\citenamefont{Mestrom
  et~al.}(2019{\natexlab{a}})\citenamefont{Mestrom, Colussi, Secker, and
  Kokkelmans}}]{mestrom2019hypervolumeSqW}
\bibinfo{author}{\bibfnamefont{P.~M.~A.} \bibnamefont{Mestrom}},
  \bibinfo{author}{\bibfnamefont{V.~E.} \bibnamefont{Colussi}},
  \bibinfo{author}{\bibfnamefont{T.}~\bibnamefont{Secker}}, \bibnamefont{and}
  \bibinfo{author}{\bibfnamefont{S.~J. J. M.~F.} \bibnamefont{Kokkelmans}},
  \bibinfo{journal}{Phys. Rev. A} \textbf{\bibinfo{volume}{100}},
  \bibinfo{pages}{050702(R)} (\bibinfo{year}{2019}{\natexlab{a}}).

\bibitem[{\citenamefont{Berninger et~al.}(2011)\citenamefont{Berninger,
  Zenesini, Huang, Harm, N\"agerl, Ferlaino, Grimm, Julienne, and
  Hutson}}]{berninger2011cesium133}
\bibinfo{author}{\bibfnamefont{M.}~\bibnamefont{Berninger}},
  \bibinfo{author}{\bibfnamefont{A.}~\bibnamefont{Zenesini}},
  \bibinfo{author}{\bibfnamefont{B.}~\bibnamefont{Huang}},
  \bibinfo{author}{\bibfnamefont{W.}~\bibnamefont{Harm}},
  \bibinfo{author}{\bibfnamefont{H.-C.} \bibnamefont{N\"agerl}},
  \bibinfo{author}{\bibfnamefont{F.}~\bibnamefont{Ferlaino}},
  \bibinfo{author}{\bibfnamefont{R.}~\bibnamefont{Grimm}},
  \bibinfo{author}{\bibfnamefont{P.~S.} \bibnamefont{Julienne}},
  \bibnamefont{and} \bibinfo{author}{\bibfnamefont{J.~M.}
  \bibnamefont{Hutson}}, \bibinfo{journal}{Phys. Rev. Lett.}
  \textbf{\bibinfo{volume}{107}}, \bibinfo{pages}{120401}
  (\bibinfo{year}{2011}).

\bibitem[{\citenamefont{Naidon and Endo}(2017)}]{naidon2017review}
\bibinfo{author}{\bibfnamefont{P.}~\bibnamefont{Naidon}} \bibnamefont{and}
  \bibinfo{author}{\bibfnamefont{S.}~\bibnamefont{Endo}},
  \bibinfo{journal}{Rep. Prog. Phys.} \textbf{\bibinfo{volume}{80}},
  \bibinfo{pages}{056001} (\bibinfo{year}{2017}).

\bibitem[{\citenamefont{Greene et~al.}(2017)\citenamefont{Greene, Giannakeas,
  and P\'erez-R\'{\i}os}}]{greene2017review}
\bibinfo{author}{\bibfnamefont{C.~H.} \bibnamefont{Greene}},
  \bibinfo{author}{\bibfnamefont{P.}~\bibnamefont{Giannakeas}},
  \bibnamefont{and}
  \bibinfo{author}{\bibfnamefont{J.}~\bibnamefont{P\'erez-R\'{\i}os}},
  \bibinfo{journal}{Rev. Mod. Phys.} \textbf{\bibinfo{volume}{89}},
  \bibinfo{pages}{035006} (\bibinfo{year}{2017}).

\bibitem[{\citenamefont{D'Incao}(2018)}]{dincao2018review}
\bibinfo{author}{\bibfnamefont{J.~P.} \bibnamefont{D'Incao}},
  \bibinfo{journal}{Journal of Physics B: Atomic, Molecular and Optical
  Physics} \textbf{\bibinfo{volume}{51}}, \bibinfo{pages}{043001}
  (\bibinfo{year}{2018}).

\bibitem[{\citenamefont{Wang et~al.}(2012{\natexlab{a}})\citenamefont{Wang,
  D'Incao, Esry, and Greene}}]{wang2012origin}
\bibinfo{author}{\bibfnamefont{J.}~\bibnamefont{Wang}},
  \bibinfo{author}{\bibfnamefont{J.~P.} \bibnamefont{D'Incao}},
  \bibinfo{author}{\bibfnamefont{B.~D.} \bibnamefont{Esry}}, \bibnamefont{and}
  \bibinfo{author}{\bibfnamefont{C.~H.} \bibnamefont{Greene}},
  \bibinfo{journal}{Phys. Rev. Lett.} \textbf{\bibinfo{volume}{108}},
  \bibinfo{pages}{263001} (\bibinfo{year}{2012}{\natexlab{a}}).

\bibitem[{\citenamefont{Wang et~al.}(2012{\natexlab{b}})\citenamefont{Wang,
  Wang, D'Incao, and Greene}}]{dincao2012heteronuclear}
\bibinfo{author}{\bibfnamefont{Y.}~\bibnamefont{Wang}},
  \bibinfo{author}{\bibfnamefont{J.}~\bibnamefont{Wang}},
  \bibinfo{author}{\bibfnamefont{J.~P.} \bibnamefont{D'Incao}},
  \bibnamefont{and} \bibinfo{author}{\bibfnamefont{C.~H.}
  \bibnamefont{Greene}}, \bibinfo{journal}{Phys. Rev. Lett.}
  \textbf{\bibinfo{volume}{109}}, \bibinfo{pages}{243201}
  (\bibinfo{year}{2012}{\natexlab{b}}).

\bibitem[{\citenamefont{Schmidt et~al.}(2012)\citenamefont{Schmidt, Rath, and
  Zwerger}}]{schmidt2012universality}
\bibinfo{author}{\bibfnamefont{R.}~\bibnamefont{Schmidt}},
  \bibinfo{author}{\bibfnamefont{S.}~\bibnamefont{Rath}}, \bibnamefont{and}
  \bibinfo{author}{\bibfnamefont{W.}~\bibnamefont{Zwerger}},
  \bibinfo{journal}{The European Physical Journal B}
  \textbf{\bibinfo{volume}{85}}, \bibinfo{pages}{386} (\bibinfo{year}{2012}).

\bibitem[{\citenamefont{Naidon et~al.}(2014)\citenamefont{Naidon, Endo, and
  Ueda}}]{naidon2014physicalorigin}
\bibinfo{author}{\bibfnamefont{P.}~\bibnamefont{Naidon}},
  \bibinfo{author}{\bibfnamefont{S.}~\bibnamefont{Endo}}, \bibnamefont{and}
  \bibinfo{author}{\bibfnamefont{M.}~\bibnamefont{Ueda}},
  \bibinfo{journal}{Phys. Rev. A} \textbf{\bibinfo{volume}{90}},
  \bibinfo{pages}{022106} (\bibinfo{year}{2014}).

\bibitem[{\citenamefont{Blume}(2015)}]{blume2015TBPhelium}
\bibinfo{author}{\bibfnamefont{D.}~\bibnamefont{Blume}},
  \bibinfo{journal}{Few-Body Systems} \textbf{\bibinfo{volume}{56}},
  \bibinfo{pages}{859} (\bibinfo{year}{2015}).

\bibitem[{\citenamefont{Naidon et~al.}(2012)\citenamefont{Naidon, Hiyama, and
  Ueda}}]{naidon2012threebodyparameter4He}
\bibinfo{author}{\bibfnamefont{P.}~\bibnamefont{Naidon}},
  \bibinfo{author}{\bibfnamefont{E.}~\bibnamefont{Hiyama}}, \bibnamefont{and}
  \bibinfo{author}{\bibfnamefont{M.}~\bibnamefont{Ueda}},
  \bibinfo{journal}{Phys. Rev. A} \textbf{\bibinfo{volume}{86}},
  \bibinfo{pages}{012502} (\bibinfo{year}{2012}).

\bibitem[{\citenamefont{S\o{}rensen et~al.}(2013)\citenamefont{S\o{}rensen,
  Fedorov, Jensen, and Zinner}}]{sorensen2013recombinationoptical}
\bibinfo{author}{\bibfnamefont{P.~K.} \bibnamefont{S\o{}rensen}},
  \bibinfo{author}{\bibfnamefont{D.~V.} \bibnamefont{Fedorov}},
  \bibinfo{author}{\bibfnamefont{A.~S.} \bibnamefont{Jensen}},
  \bibnamefont{and} \bibinfo{author}{\bibfnamefont{N.~T.}
  \bibnamefont{Zinner}}, \bibinfo{journal}{Phys. Rev. A}
  \textbf{\bibinfo{volume}{88}}, \bibinfo{pages}{042518}
  (\bibinfo{year}{2013}).

\bibitem[{\citenamefont{Chapurin et~al.}(2019)\citenamefont{Chapurin, Xie,
  Van~de Graaff, Popowski, D'Incao, Julienne, Ye, and
  Cornell}}]{chapurin2019precision}
\bibinfo{author}{\bibfnamefont{R.}~\bibnamefont{Chapurin}},
  \bibinfo{author}{\bibfnamefont{X.}~\bibnamefont{Xie}},
  \bibinfo{author}{\bibfnamefont{M.~J.} \bibnamefont{Van~de Graaff}},
  \bibinfo{author}{\bibfnamefont{J.~S.} \bibnamefont{Popowski}},
  \bibinfo{author}{\bibfnamefont{J.~P.} \bibnamefont{D'Incao}},
  \bibinfo{author}{\bibfnamefont{P.~S.} \bibnamefont{Julienne}},
  \bibinfo{author}{\bibfnamefont{J.}~\bibnamefont{Ye}}, \bibnamefont{and}
  \bibinfo{author}{\bibfnamefont{E.~A.} \bibnamefont{Cornell}},
  \bibinfo{journal}{Phys. Rev. Lett.} \textbf{\bibinfo{volume}{123}},
  \bibinfo{pages}{233402} (\bibinfo{year}{2019}).

\bibitem[{\citenamefont{Alt et~al.}(1967)\citenamefont{Alt, Grassberger, and
  Sandhas}}]{alt1967ags}
\bibinfo{author}{\bibfnamefont{E.}~\bibnamefont{Alt}},
  \bibinfo{author}{\bibfnamefont{P.}~\bibnamefont{Grassberger}},
  \bibnamefont{and} \bibinfo{author}{\bibfnamefont{W.}~\bibnamefont{Sandhas}},
  \bibinfo{journal}{Nucl. Phys. B} \textbf{\bibinfo{volume}{2}},
  \bibinfo{pages}{167 } (\bibinfo{year}{1967}).

\bibitem[{\citenamefont{Weinberg}(1963)}]{weinberg1963expansion}
\bibinfo{author}{\bibfnamefont{S.}~\bibnamefont{Weinberg}},
  \bibinfo{journal}{Phys. Rev.} \textbf{\bibinfo{volume}{131}},
  \bibinfo{pages}{440} (\bibinfo{year}{1963}).

\bibitem[{\citenamefont{Mestrom
  et~al.}(2019{\natexlab{b}})\citenamefont{Mestrom, Secker, Kroeze, and
  Kokkelmans}}]{mestrom2019squarewell}
\bibinfo{author}{\bibfnamefont{P.~M.~A.} \bibnamefont{Mestrom}},
  \bibinfo{author}{\bibfnamefont{T.}~\bibnamefont{Secker}},
  \bibinfo{author}{\bibfnamefont{R.~M.} \bibnamefont{Kroeze}},
  \bibnamefont{and} \bibinfo{author}{\bibfnamefont{S.~J. J. M.~F.}
  \bibnamefont{Kokkelmans}}, \bibinfo{journal}{Phys. Rev. A}
  \textbf{\bibinfo{volume}{99}}, \bibinfo{pages}{012702}
  (\bibinfo{year}{2019}{\natexlab{b}}).

\bibitem[{Sup()}]{SupplMat}
\bibinfo{note}{See Supplemental Material for additional details of our
  calculations and results.}

\bibitem[{not({\natexlab{a}})}]{noteBackgroundScatteringLength}
\bibinfo{note}{The scattering length contains a resonant and nonresonant
  contribution. The latter is known as the background scattering length and is
  on the order of $r_{\mathrm{vdW}}$ for the considered van der Waals
  potentials.}

\bibitem[{not({\natexlab{b}})}]{noteSharpTrimerResonances}
\bibinfo{note}{There could be additional narrow trimer resonances for which our
  grid for $a$ is too sparse. Such resonances would have a width smaller than
  $\delta a = 10^{-3}~r_{\mathrm{vdW}}$.}

\bibitem[{not({\natexlab{c}})}]{noteResultsSCvdW}
\bibinfo{note}{This expectation is consistent with our results of
  $\mathrm{Re}(D)$ for the potentials $V_{\mathrm{sc}}^{(1)}$,
  $V_{\mathrm{sc}}^{(2)}$ and $V_{\mathrm{sc}}^{(3)}$ which approach the
  universal curve as the potential depth increases \cite{SupplMat}.}

\bibitem[{not({\natexlab{d}})}]{noteZwergerPrediction}
\bibinfo{note}{The uncertainty in this prediction is determined by the validity
  of the approximations that are made in Ref.~\cite{miller1977phasetransition},
  including the trial wave function that is used to describe the ground state
  of the Bose fluid near the quantum tricritical point.}

\bibitem[{\citenamefont{Miller et~al.}(1977)\citenamefont{Miller, Nosanow, and
  Parish}}]{miller1977phasetransition}
\bibinfo{author}{\bibfnamefont{M.~D.} \bibnamefont{Miller}},
  \bibinfo{author}{\bibfnamefont{L.~H.} \bibnamefont{Nosanow}},
  \bibnamefont{and} \bibinfo{author}{\bibfnamefont{L.~J.}
  \bibnamefont{Parish}}, \bibinfo{journal}{Phys. Rev. B}
  \textbf{\bibinfo{volume}{15}}, \bibinfo{pages}{214} (\bibinfo{year}{1977}).

\bibitem[{\citenamefont{Fu et~al.}(2003)\citenamefont{Fu, Wang, and
  Gao}}]{gao2003extendedGP}
\bibinfo{author}{\bibfnamefont{H.}~\bibnamefont{Fu}},
  \bibinfo{author}{\bibfnamefont{Y.}~\bibnamefont{Wang}}, \bibnamefont{and}
  \bibinfo{author}{\bibfnamefont{B.}~\bibnamefont{Gao}},
  \bibinfo{journal}{Phys. Rev. A} \textbf{\bibinfo{volume}{67}},
  \bibinfo{pages}{053612} (\bibinfo{year}{2003}).

\bibitem[{\citenamefont{Collin et~al.}(2007)\citenamefont{Collin, Massignan,
  and Pethick}}]{collin2007extendedGP}
\bibinfo{author}{\bibfnamefont{A.}~\bibnamefont{Collin}},
  \bibinfo{author}{\bibfnamefont{P.}~\bibnamefont{Massignan}},
  \bibnamefont{and} \bibinfo{author}{\bibfnamefont{C.~J.}
  \bibnamefont{Pethick}}, \bibinfo{journal}{Phys. Rev. A}
  \textbf{\bibinfo{volume}{75}}, \bibinfo{pages}{013615}
  (\bibinfo{year}{2007}).

\bibitem[{\citenamefont{Th\o{}gersen et~al.}(2009)\citenamefont{Th\o{}gersen,
  Zinner, and Jensen}}]{thogersen2009extendedGP}
\bibinfo{author}{\bibfnamefont{M.}~\bibnamefont{Th\o{}gersen}},
  \bibinfo{author}{\bibfnamefont{N.~T.} \bibnamefont{Zinner}},
  \bibnamefont{and} \bibinfo{author}{\bibfnamefont{A.~S.}
  \bibnamefont{Jensen}}, \bibinfo{journal}{Phys. Rev. A}
  \textbf{\bibinfo{volume}{80}}, \bibinfo{pages}{043625}
  (\bibinfo{year}{2009}).

\bibitem[{\citenamefont{Dalfovo et~al.}(1999)\citenamefont{Dalfovo, Giorgini,
  Pitaevskii, and Stringari}}]{RevModPhys.71.463}
\bibinfo{author}{\bibfnamefont{F.}~\bibnamefont{Dalfovo}},
  \bibinfo{author}{\bibfnamefont{S.}~\bibnamefont{Giorgini}},
  \bibinfo{author}{\bibfnamefont{L.~P.} \bibnamefont{Pitaevskii}},
  \bibnamefont{and}
  \bibinfo{author}{\bibfnamefont{S.}~\bibnamefont{Stringari}},
  \bibinfo{journal}{Rev. Mod. Phys.} \textbf{\bibinfo{volume}{71}},
  \bibinfo{pages}{463} (\bibinfo{year}{1999}).

\bibitem[{\citenamefont{P\'erez-Garc\'{\i}a
  et~al.}(1996)\citenamefont{P\'erez-Garc\'{\i}a, Michinel, Cirac, Lewenstein,
  and Zoller}}]{PhysRevLett.77.5320}
\bibinfo{author}{\bibfnamefont{V.~M.} \bibnamefont{P\'erez-Garc\'{\i}a}},
  \bibinfo{author}{\bibfnamefont{H.}~\bibnamefont{Michinel}},
  \bibinfo{author}{\bibfnamefont{J.~I.} \bibnamefont{Cirac}},
  \bibinfo{author}{\bibfnamefont{M.}~\bibnamefont{Lewenstein}},
  \bibnamefont{and} \bibinfo{author}{\bibfnamefont{P.}~\bibnamefont{Zoller}},
  \bibinfo{journal}{Phys. Rev. Lett.} \textbf{\bibinfo{volume}{77}},
  \bibinfo{pages}{5320} (\bibinfo{year}{1996}).

\bibitem[{\citenamefont{Yi and You}(2001)}]{PhysRevA.63.053607}
\bibinfo{author}{\bibfnamefont{S.}~\bibnamefont{Yi}} \bibnamefont{and}
  \bibinfo{author}{\bibfnamefont{L.}~\bibnamefont{You}},
  \bibinfo{journal}{Phys. Rev. A} \textbf{\bibinfo{volume}{63}},
  \bibinfo{pages}{053607} (\bibinfo{year}{2001}).

\bibitem[{\citenamefont{Al-Jibbouri et~al.}(2013)\citenamefont{Al-Jibbouri,
  Vidanovi{\'{c}}, Bala{\v{z}}, and Pelster}}]{Al_Jibbouri_2013}
\bibinfo{author}{\bibfnamefont{H.}~\bibnamefont{Al-Jibbouri}},
  \bibinfo{author}{\bibfnamefont{I.}~\bibnamefont{Vidanovi{\'{c}}}},
  \bibinfo{author}{\bibfnamefont{A.}~\bibnamefont{Bala{\v{z}}}},
  \bibnamefont{and} \bibinfo{author}{\bibfnamefont{A.}~\bibnamefont{Pelster}},
  \bibinfo{journal}{Journal of Physics B: Atomic, Molecular and Optical
  Physics} \textbf{\bibinfo{volume}{46}}, \bibinfo{pages}{065303}
  (\bibinfo{year}{2013}).

\bibitem[{\citenamefont{Kohn}(1961)}]{PhysRev.123.1242}
\bibinfo{author}{\bibfnamefont{W.}~\bibnamefont{Kohn}}, \bibinfo{journal}{Phys.
  Rev.} \textbf{\bibinfo{volume}{123}}, \bibinfo{pages}{1242}
  (\bibinfo{year}{1961}).

\bibitem[{\citenamefont{Stringari}(1996)}]{PhysRevLett.77.2360}
\bibinfo{author}{\bibfnamefont{S.}~\bibnamefont{Stringari}},
  \bibinfo{journal}{Phys. Rev. Lett.} \textbf{\bibinfo{volume}{77}},
  \bibinfo{pages}{2360} (\bibinfo{year}{1996}).

\bibitem[{\citenamefont{Straatsma et~al.}(2016)\citenamefont{Straatsma,
  Colussi, Davis, Lobser, Holland, Anderson, Lewandowski, and
  Cornell}}]{PhysRevA.94.043640}
\bibinfo{author}{\bibfnamefont{C.~J.~E.} \bibnamefont{Straatsma}},
  \bibinfo{author}{\bibfnamefont{V.~E.} \bibnamefont{Colussi}},
  \bibinfo{author}{\bibfnamefont{M.~J.} \bibnamefont{Davis}},
  \bibinfo{author}{\bibfnamefont{D.~S.} \bibnamefont{Lobser}},
  \bibinfo{author}{\bibfnamefont{M.~J.} \bibnamefont{Holland}},
  \bibinfo{author}{\bibfnamefont{D.~Z.} \bibnamefont{Anderson}},
  \bibinfo{author}{\bibfnamefont{H.~J.} \bibnamefont{Lewandowski}},
  \bibnamefont{and} \bibinfo{author}{\bibfnamefont{E.~A.}
  \bibnamefont{Cornell}}, \bibinfo{journal}{Phys. Rev. A}
  \textbf{\bibinfo{volume}{94}}, \bibinfo{pages}{043640}
  (\bibinfo{year}{2016}).

\bibitem[{\citenamefont{Shotan et~al.}(2014)\citenamefont{Shotan, Machtey,
  Kokkelmans, and Khaykovich}}]{khaykovich2014zeroa0}
\bibinfo{author}{\bibfnamefont{Z.}~\bibnamefont{Shotan}},
  \bibinfo{author}{\bibfnamefont{O.}~\bibnamefont{Machtey}},
  \bibinfo{author}{\bibfnamefont{S.}~\bibnamefont{Kokkelmans}},
  \bibnamefont{and}
  \bibinfo{author}{\bibfnamefont{L.}~\bibnamefont{Khaykovich}},
  \bibinfo{journal}{Phys. Rev. Lett.} \textbf{\bibinfo{volume}{113}},
  \bibinfo{pages}{053202} (\bibinfo{year}{2014}).

\bibitem[{\citenamefont{Pricoupenko and
  Petrov}(2019)}]{petrov2019zeroScatL3bodyInteraction}
\bibinfo{author}{\bibfnamefont{A.}~\bibnamefont{Pricoupenko}} \bibnamefont{and}
  \bibinfo{author}{\bibfnamefont{D.~S.} \bibnamefont{Petrov}},
  \bibinfo{journal}{Phys. Rev. A} \textbf{\bibinfo{volume}{100}},
  \bibinfo{pages}{042707} (\bibinfo{year}{2019}).

\bibitem[{\citenamefont{Chin et~al.}(2010)\citenamefont{Chin, Grimm, Julienne,
  and Tiesinga}}]{chin2010feshbach}
\bibinfo{author}{\bibfnamefont{C.}~\bibnamefont{Chin}},
  \bibinfo{author}{\bibfnamefont{R.}~\bibnamefont{Grimm}},
  \bibinfo{author}{\bibfnamefont{P.}~\bibnamefont{Julienne}}, \bibnamefont{and}
  \bibinfo{author}{\bibfnamefont{E.}~\bibnamefont{Tiesinga}},
  \bibinfo{journal}{Rev. Mod. Phys.} \textbf{\bibinfo{volume}{82}},
  \bibinfo{pages}{1225} (\bibinfo{year}{2010}).

\end{thebibliography}

%%%%%%%%%%%%%%%% SUPPLEMENTARY
\clearpage
\pagebreak

\onecolumngrid
\begin{center}
  \textbf{\large Supplemental Material: ``Van der Waals universality near a quantum tricritical point''}\\[.2cm]
  P. M. A. Mestrom,$^{1}$ V. E. Colussi,$^{1}$ T. Secker,$^{1}$ G. P. Groeneveld,$^{1}$ and S. J. J. M. F. Kokkelmans$^1$\\[.1cm]
  {\itshape ${}^1$Eindhoven University of Technology, P.~O.~Box 513, 5600 MB Eindhoven, The Netherlands\\}
%(Dated: \today)\\[1cm]
\end{center}
\twocolumngrid

\setcounter{equation}{0}
\setcounter{figure}{0}
\setcounter{page}{1}
\renewcommand{\theequation}{S\arabic{equation}}
\renewcommand{\thefigure}{S\arabic{figure}}
\renewcommand{\thetable}{S\arabic{table}}  
\renewcommand{\bibnumfmt}[1]{[S#1]}

\section{Two-body interaction models}

\begin{figure}[hbtp]
    \centering
    \includegraphics[width=3.4in]{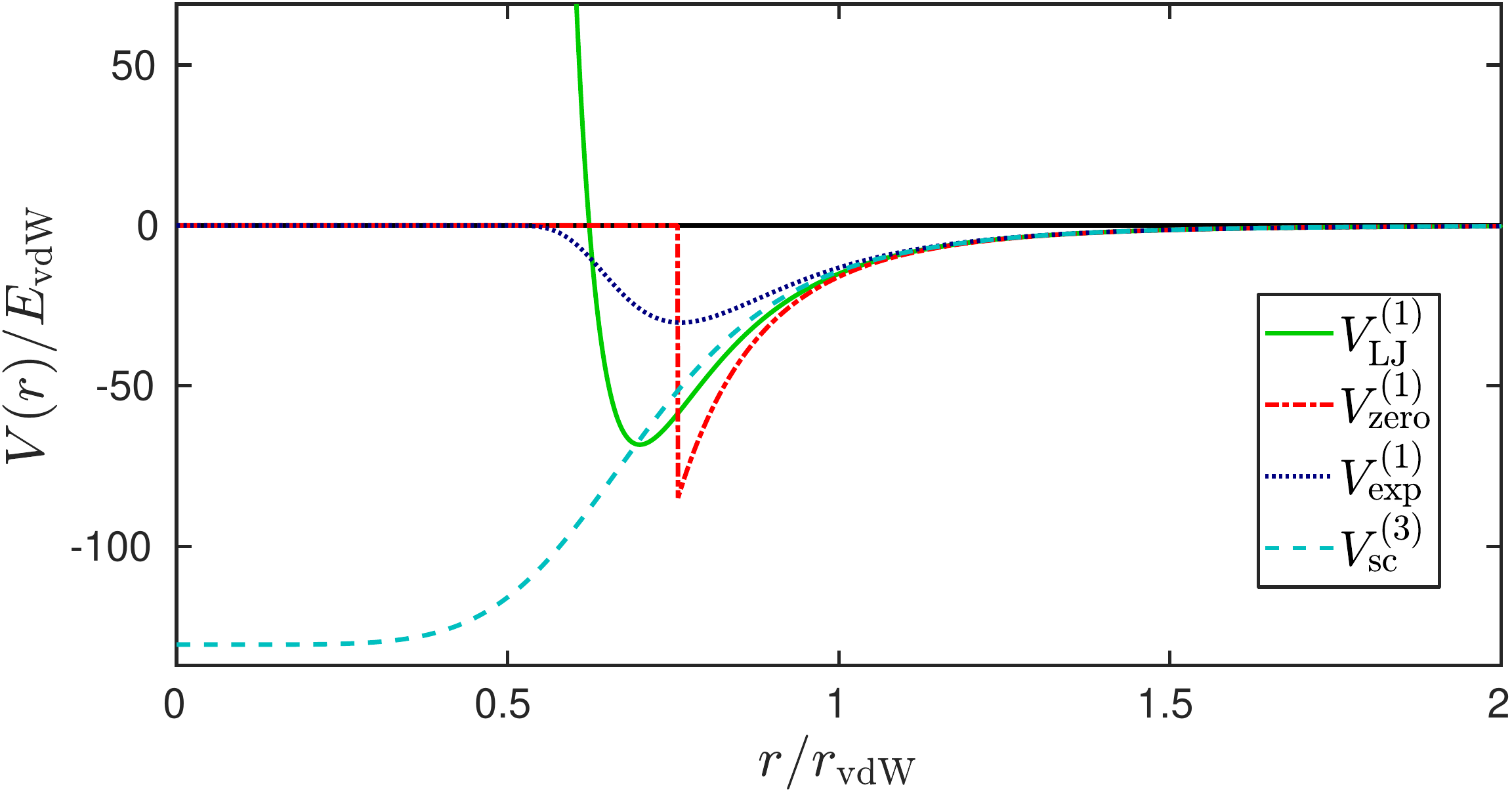}
    \caption{The considered van der Waals potentials in units of $E_{\mathrm{vdW}}~\equiv~\hbar^2/(m r_{\mathrm{vdW}}^2)$ as a function of the interparticle distance $r$. The potential depths are chosen such that $a = 0$.}
    \label{fig:vdW_potentials}
\end{figure}

In this work, we have analyzed the behavior of the scattering hypervolume using the following interaction potentials to mimic interatomic interactions:
\begin{align}
V_{\mathrm{LJ}} &= -\frac{C_6}{r^6}\left(1-\frac{\lambda^6}{r^6}\right),\label{eq:V_LJ}
\\
V_{\mathrm{zero}} &= -\frac{C_6}{r^6} \Theta(r -\lambda), \label{eq:V_zero}
\\
V_{\mathrm{exp}} &= -\frac{C_6}{r^6} \exp(-\lambda^6/r^6), \label{eq:V_exp}
\\
V_{\mathrm{sc}} &= -\frac{C_6}{r^6 + \lambda^6}. \label{eq:V_sc}
\end{align}
These potentials are displayed in \refFigure{fig:vdW_potentials}. They are all dominated by the van der Waals interaction, $-C_6/r^6$, for large interparticle separations $r$ (i.e., $r\gg \lambda$). The potential in \refEquation{eq:V_LJ} is known as the Lennard-Jones potential. It is strongly repulsive for $r < \lambda$. To handle the infinitely strong repulsive barrier numerically, we cut off this barrier as follows:
\begin{equation}
V(r)=\begin{cases}
-\frac{C_6}{r_0^6}\left(1-\frac{\lambda^6}{r_0^6}\right),&\mbox{$0\leq r<r_0$},
\\
-\frac{C_6}{r^6}\left(1-\frac{\lambda^6}{r^6}\right),&\mbox{$r\geq r_0$}.
\end{cases}
\end{equation}
We choose $r_0$ small enough such that the height of the barrier is much larger than the depth of the well. In \refEquation{eq:V_zero}, $\Theta(x)$ is the Heaviside step function. It is zero for $x<0$ and $1$ for $x\geq 0$. This potential is thus zero for $r<\lambda$ and consists of a pure van der Waals tail for $r\geq\lambda$. The potential $V_{\mathrm{exp}}$ defined in \refEquation{eq:V_exp} also goes to zero for $r \to 0$, but in a smooth way. Finally, \refEquation{eq:V_sc} represents a van der Waals potential with a ``soft core". It is attractive for all values of $r$.

% Figure made using the code:  Plot_Vtotal_v3_LJvdW_ExpvdW_ZerovdW_SCvdW_v2.m

\section{Three-body resonances}

\begin{figure}[hbtp]
    \centering
    \includegraphics[width=3.4in]{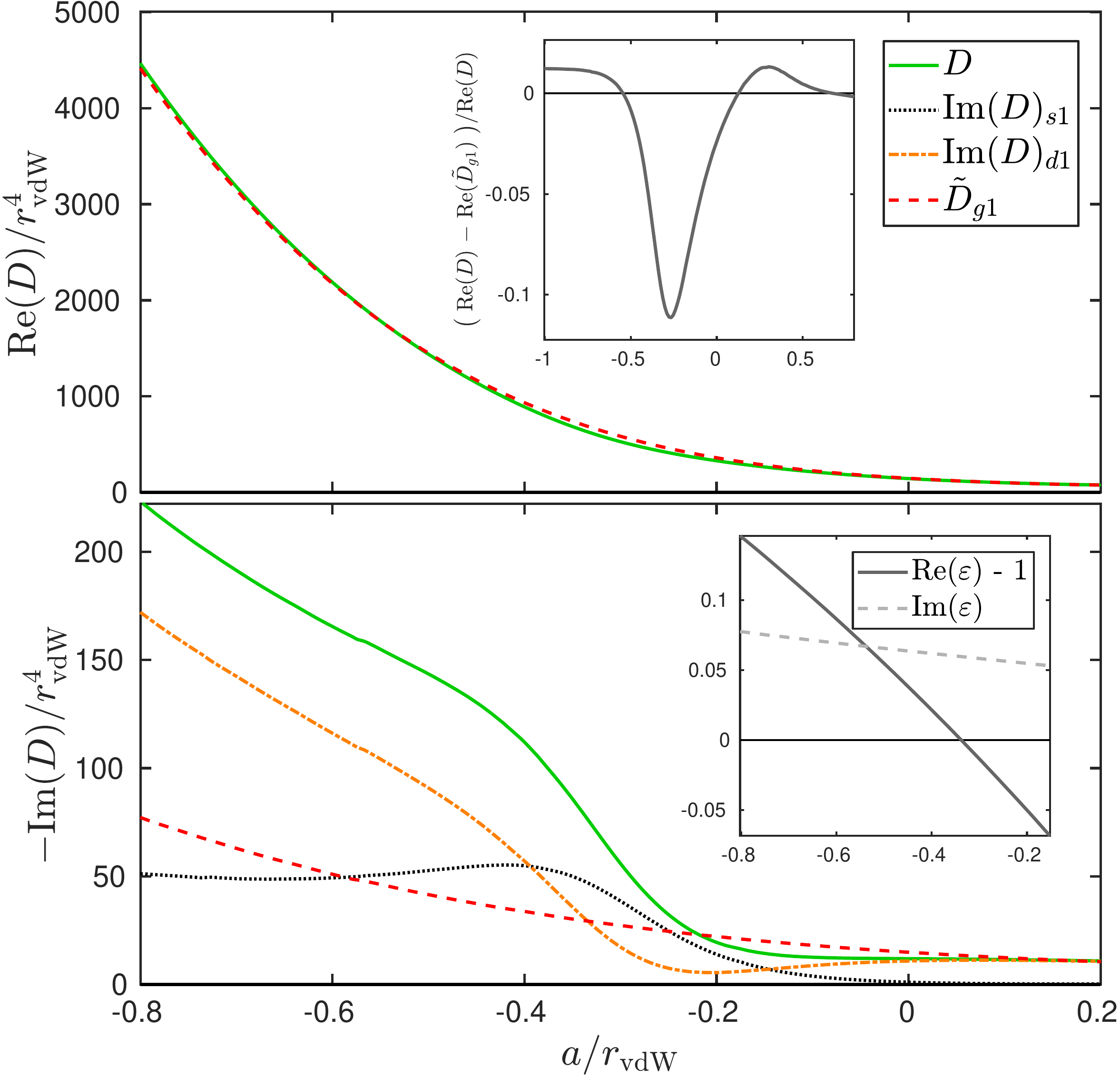}
    \caption{Three-body scattering hypervolume $D$ corresponding to the Lennard-Jones potential $V_{\mathrm{LJ}}^{(1)}$ as a function of the two-body scattering length $a$. Its imaginary part can be written as $\mathrm{Im}(D) = \mathrm{Im}(D)_{s1} + \mathrm{Im}(D)_{d1}$, where $\mathrm{Im}(D)_{s1}$ and $\mathrm{Im}(D)_{d1}$ are proportional to the partial three-body recombination rates into the first $s$-wave and $d$-wave dimer state, respectively. Additionally, $\tilde{D}_{g1}$ is the scattering hypervolume obtained by excluding the first $g$-wave term of the Weinberg expansion.  The upper inset shows the relative difference between the real parts of $D$ and $\tilde{D}_{g1}$. The lower inset shows the behavior of the eigenvalue $\varepsilon$ of the kernel of the three-body intergal equation causing this trimer resonance. Near the resonance position, the real part of $\varepsilon$ passes one as discussed in Ref.~\cite{mestrom2019squarewell}.}
    \label{fig:LJvdW_Res1and2_Dhyp_gwave_trimer}
\end{figure}

In the weakly interacting regime, the scattering hypervolume is influenced by nonuniversal trimer resonances \cite{mestrom2019hypervolumeSqW, tan2017hypervolume}. Figure~1 of the main text shows several of such trimer resonances. Some resonances have only a weak effect on $D$. For example, the Lennard-Jones potential $V_{\mathrm{LJ}}^{(1)}$ supports a three-body quasibound state at zero energy near $a/r_{\mathrm{vdW}}\simeq -0.3$. Figure~\ref{fig:LJvdW_Res1and2_Dhyp_gwave_trimer} shows that this trimer resonance vanishes when the two-body $g$-wave state that gets bound at $a/r_{\mathrm{vdW}} = -4.8$ is removed from the Weinberg expansion of the two-body interaction potential $V$. This demonstrates its strong $g$-wave character.

% Figure made with the code: Plot_and_fit_Elastic_K3_List_v27_LJvdW_Res1and2_a0_0d2_m0d8_eig.m

%\section{Hard-hypersphere collisions}

%In \del{Fig.~2 of the main text}, we have shown that the hard-hypersphere radius $R_{\mathrm{hh}}$ is well described by $|a - a_{\mathrm{hh}}^{+}|$ and $|a - a_{\mathrm{hh}}^{-}|$ in the regimes $0.6 \lesssim a/r_{\mathrm{vdW}}\lesssim 1$ and $-1 \lesssim a/r_{\mathrm{vdW}}\lesssim -0.1$, respectively. Figure~\ref{fig:LJvdW_hard_sphere_radius_SuppMat} shows the same result for the real part of the three-body scattering hypervolume.

%\begin{figure}[hbtp]
%    \centering
%    \includegraphics[width=3.4in]{Figure_LJvdW_Dhyp_PotRes_1and2_a0_-1_1_Dhyp_0_8500_Comparison_HS_a0shiftPlus_-0d01243_a0shiftMinus_0d47757_v3.pdf}
%    \caption{Real part of the three-body scattering hypervolume $D$ corresponding to the potential $V_{\mathrm{LJ}}^{(1)}$ (green solid line) as a function of the two-body scattering length $a$. The dashed and dash-dotted curves show the hard-hypersphere formula for which the hard-hypersphere radius is modified to match $\mathrm{Re}(D)$ at $a/r_{\mathrm{vdW}} = 1$ and $a/r_{\mathrm{vdW}} = -1$, respectively. The values of $a_{\mathrm{hh}}^{+}/r_{\mathrm{vdW}}$ and $a_{\mathrm{hh}}^{-}/r_{\mathrm{vdW}}$ are $-0.012$ and $0.477$, respectively.}
%    \label{fig:LJvdW_hard_sphere_radius_SuppMat}
%\end{figure}
% Figure made using the code: Plot_and_fit_Elastic_K3_List_v30_LJvdW_scaling_a0shift_v2.m

\section{Additional results for the soft-core van der Waals potential}

\begin{figure}[htbp] %[hb!]
    \centering
    \includegraphics[width=3.4in]{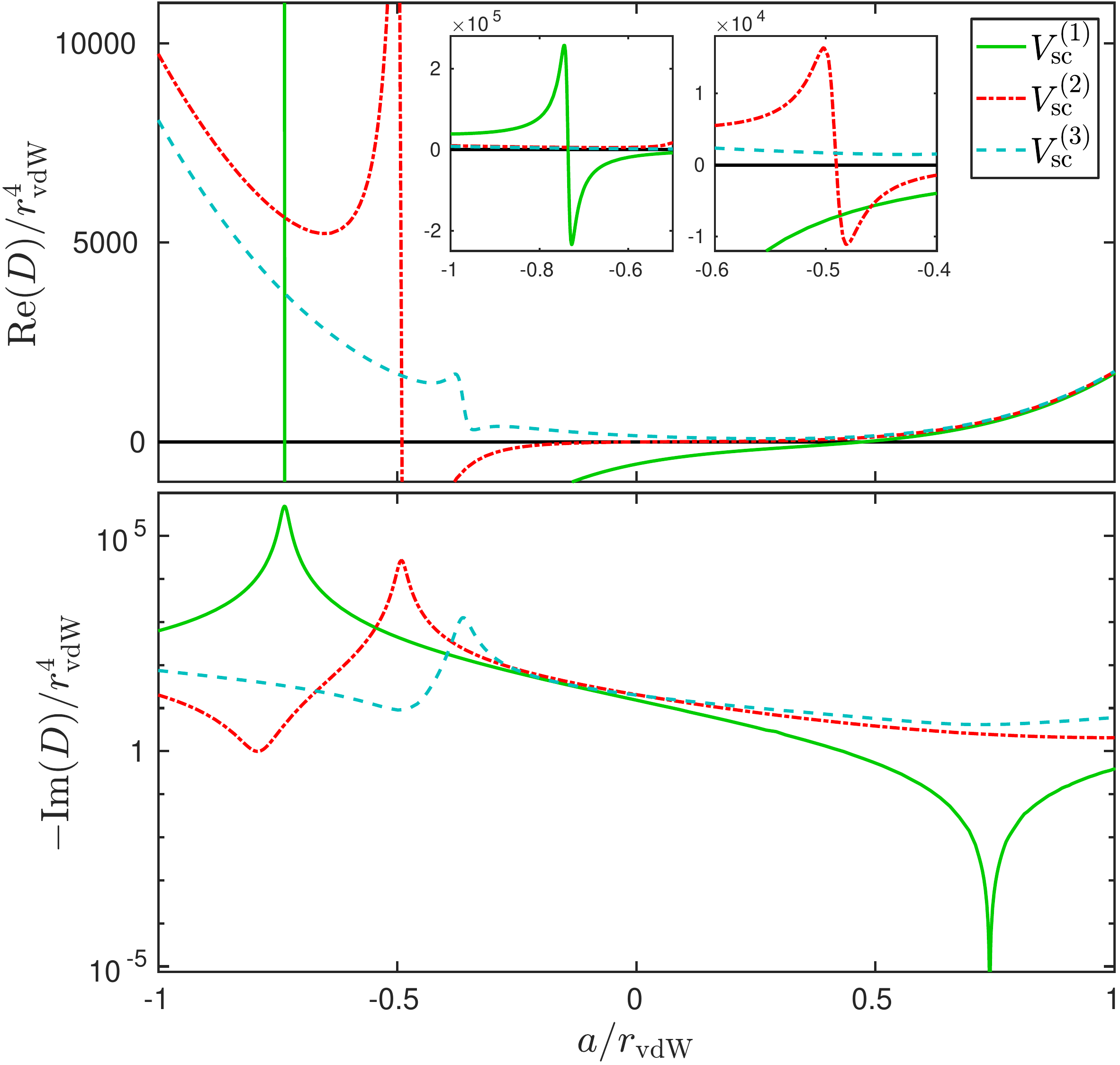}
    \caption{Three-body scattering hypervolume $D$ corresponding to soft-core van der Waals potentials supporting one, two and three $s$-wave dimer states as a function of the two-body scattering length $a$. The insets display some trimer resonances.}
    \label{fig:SCvdW_Dh_MultiplePotentials}
\end{figure}

We have calculated the three-body scattering hypervolume $D$ in the weakly interacting regime corresponding to the soft-core van der Waals potential supporting one, two and three $s$-wave dimer states. These results are presented in \refFigure{fig:SCvdW_Dh_MultiplePotentials}. This figure shows that the real part of $D$ in the weakly interacting regime for $V_{\mathrm{sc}}^{(1)}$ and $V_{\mathrm{sc}}^{(2)}$ is strongly influenced by trimer resonances and deviates strongly from $\mathrm{Re}(D)$ for $V_{\mathrm{sc}}^{(3)}$. Therefore, we conclude that the potentials $V_{\mathrm{sc}}^{(1)}$ and $V_{\mathrm{sc}}^{(2)}$ are too shallow to obey the van der Waals universality in $\mathrm{Re}(D)$. Furthermore, we find a sharp minimum in $-\mathrm{Im}(D)$ for $V_{\mathrm{sc}}^{(1)}$ at $a/r_{\mathrm{vdW}} = 0.739$ which is likely to be caused by destructive interference between competing pathways for three-body recombination \cite{dincao2018review}. 

% Figure made with the code Plot_and_fit_Elastic_K3_List_v30_SCvdW_1_2_a_2_3_a_3_4_v4b.m

\section{Additional results for the square-well potential}

\begin{figure}[htbp] %[hb!]
    \centering
    \includegraphics[width=3.4in]{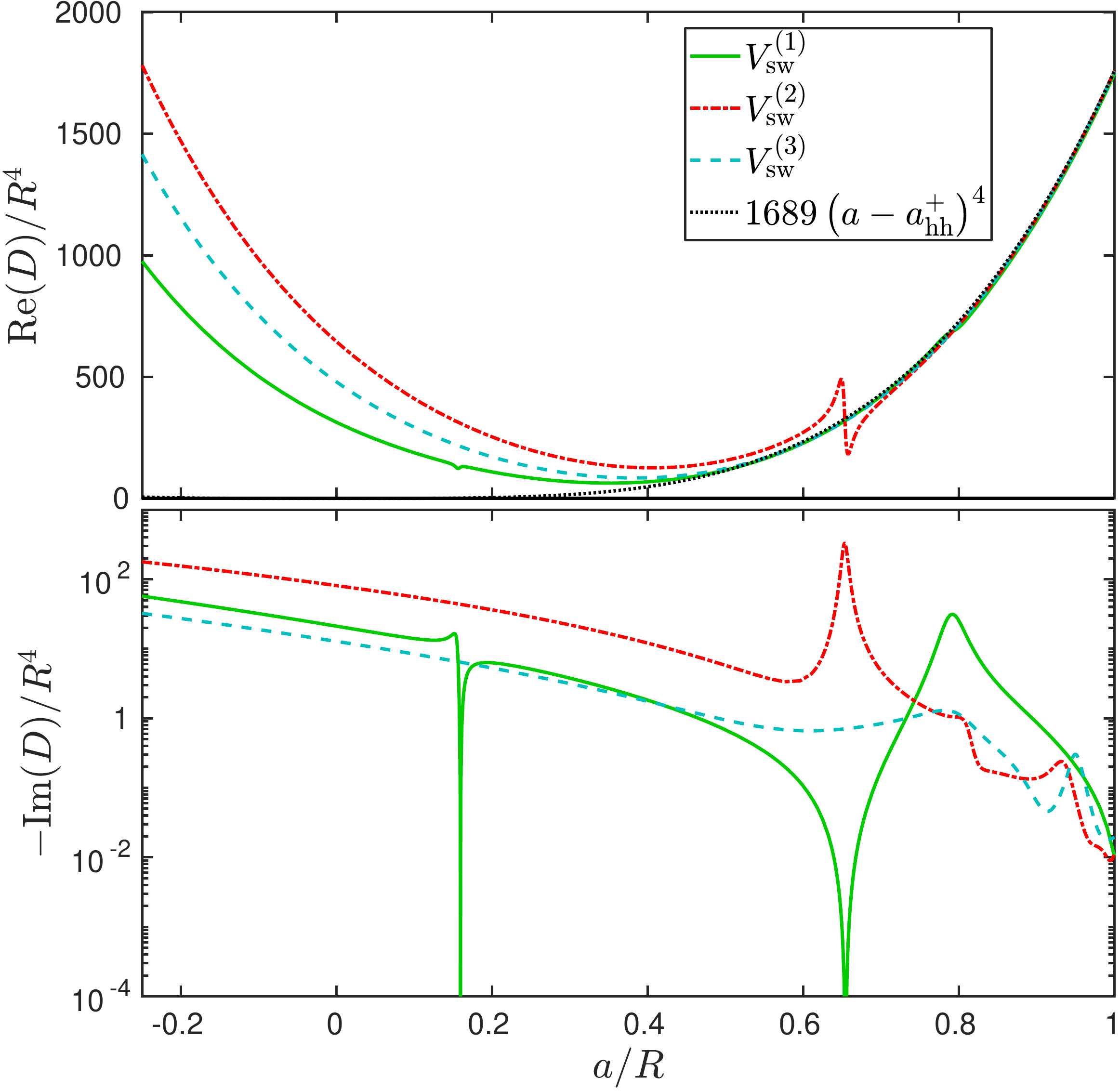}
    \caption{Three-body scattering hypervolume $D$ corresponding to square-well potentials supporting one, two and three $s$-wave dimer states as a function of the two-body scattering length $a$. We have presented the results for $V_{\mathrm{sw}}^{(1)}$ and $V_{\mathrm{sw}}^{(2)}$ before in Ref.~\cite{mestrom2019hypervolumeSqW}. The black dotted line corresponds to the curve $D = 1689\,(a - a_{\mathrm{hh}}^{+})^4$ where $a_{\mathrm{hh}}^{+}/R = -0.01$. The choice of this value is based on the values of $D$ for $V_{\mathrm{sw}}^{(1)}$, $V_{\mathrm{sw}}^{(2)}$ and $V_{\mathrm{sw}}^{(3)}$ at $a/R = 1$.}
    \label{fig:SqW_Dh_MultiplePotentials}
\end{figure}

In this section we consider identical bosons interacting via pairwise square-well potentials defined by
\begin{equation}
V_{\mathrm{sw}}(r)=\begin{cases}-V_0,&\mbox{$0\leq r<R$},\\0,&\mbox{$r\geq R$}.\end{cases}
\end{equation}
Here $R$ and $V_0$ represent the range and depth of the potential, respectively.
We have calculated the three-body scattering hypervolume $D$ in the weakly interacting regime corresponding to the square-well potential supporting one, two and three $s$-wave dimer states. Parts of these results have been presented in Ref.~\cite{mestrom2019hypervolumeSqW}. Figure~\ref{fig:SqW_Dh_MultiplePotentials} shows the results as a function of the two-body scattering length $a$. Clearly, the real part of $D$ still follows the universal $a^4$ scaling for $a/R \gtrsim 0.7$. However, this universality vanishes as $a$ decreases.

% Figure made with the code Plot_and_fit_Elastic_K3_List_v30_SqW_Res_1_2_and_2_3_and_3_4_v6.m

\section{Zwerger's prediction}

Here we consider a Bose fluid at zero temperature and zero presure. The bosons interact via a pairwise Lennard-Jones potential (see \refEquation{eq:V_LJ}) that supports no dimer states ($V_{\mathrm{LJ}}^{(0)}$).
According to Ref.~\cite{zwerger2019phasetransition}, the scattering hypervolume at $a = 0$ is related to the curvature $\tilde{u}_c^{''}$ of the energy per particle of this Bose fluid by
\begin{equation}\label{eq:Dh_prediction_Zwerger}
D = -3 \pi^2 a_{\Lambda}^2 \lambda^2 r_{\mathrm{vdW}}^2 \frac{1}{\tilde{u}_c^{''}}.
\end{equation}
Here $a_{\Lambda}$ is a dimensionless parameter that characterizes the zero crossing of $a$ \cite{zwerger2019phasetransition}. From the scattering length of $V_{\mathrm{LJ}}^{(0)}$, we find that $a_{\Lambda} = 3.82812(5)$ and $r_{\mathrm{vdW}}/\lambda = 0.858036(1)$ at $a = 0$. Since $\tilde{u}_c^{''} = -6.547$ \cite{miller1977phasetransition}, \refEquation{eq:Dh_prediction_Zwerger} gives $D/r_{\mathrm{vdW}}^4 = 90.02$. The uncertainty in this prediction is determined by the uncertainty in the value of $\tilde{u}_c^{''}$.

\section{Collective Modes}

Here we provide more details about the variational method used to obtain the collective mode spectrum at $a=0$, noting that similar calculations for a Gross-Pitaevskii equation with a cubic nonlinearity have appeared also in Ref.~\cite{Al_Jibbouri_2013}.  Beginning from the Gaussian trial wave function \cite{PhysRevLett.77.5320,PhysRevA.63.053607} given by Eq.~(7) of the main text
\begin{equation}%\label{eq:Gaussian_ansatz_excitations}
\Psi(x,y,z,t)= A(t) \prod_{\eta=x,y,z}e^{-\frac{(\eta-\eta_0)^2}{2w_\eta^2}+i\eta\alpha_\eta+i\eta^2\beta_\eta},
\end{equation}
 we obtain the Lagrangian
 \begin{widetext}
\begin{align}
\frac{2L}{N}=&-\left[\sum_{\eta}\left(\hbar\partial_t\beta_\eta+\frac{2\hbar^2\beta^2_\eta}{m}+\frac{1}{2}m\lambda_\eta^2\omega_\mathrm{ho}^2\right)\times\left(w_\eta^2+2\eta_0^2\right)+\left(\hbar\partial_t\alpha_\eta+\frac{2\hbar^2\alpha_\eta\beta_\eta}{m}\right)2\eta_0+\frac{\hbar^2}{2mw^2_\eta}+\frac{\hbar^2\alpha^2_\eta}{m}\right]\nonumber\\
&-\frac{N^2\hbar^2D}{3^{5/2}\pi^3w_x^2w_y^2w_z^2}+i\hbar\left[\frac{\partial_tA}{A}-\frac{\partial_tA^*}{A^*}\right].
\end{align}
\end{widetext}
%\begin{align}
%\frac{2L}{N}=&i\hbar\left[\frac{\dot{A}}{A}-\frac{\dot{A}^*}{A^*}\right]-\left[\sum_{\eta}\left(\dot{\beta}_\eta w_\eta^2\hbar+\frac{\hbar^2}{2w_\eta^2 m}+\frac{2\hbar^2 w_\eta^2\beta_\eta^2}{m}\right.\right.\nonumber\\
%&+\left.\left.\frac{m\omega_\eta^2w_\eta^2}{2}\right)+\frac{N^2\hbar^2\text{Re}[D]}{3^{5/2}\pi^3w_x^2w_y^2w_z^2}\right].
%\end{align}
%VEC note to self:  Add in COM displacement and double check EOMs and remove Stringari method from main text.
From the Euler-Lagrange equations, we find 
\begin{align}
&\beta_\eta=m(\partial_t w_\eta)/2\hbar^2w_\eta,\\
&\alpha_\eta=\frac{m}{\hbar}\left(\partial_t\eta_0-\frac{\eta_0\partial_tw_\eta}{w_\eta}\right),\\
&\partial_t^2\eta_0+\lambda_\eta^2\omega_\mathrm{ho}^2\eta_0=0,\\
&\partial^2_\tau  v_{\eta}+\lambda_{\eta}^2 v_{\eta}=\frac{1}{v_{\eta}^3}+\frac{K}{v_{\eta} (v_x v_y v_z)^2},\label{eq:v}
\end{align}
with $K=2 D N^2/9\sqrt{3}\pi^3 l_{\mathrm{ho}}^4$ and dimensionless scalings $v_\eta=w_\eta/l_\mathrm{ho}$ and $\tau=\omega_\mathrm{ho} t$.  From Eq.~\eqref{eq:v}, we identify Newton's equations for classical motion in an effective potential
\begin{align}
U(v_x,v_y,v_z)&=\frac{1}{2}(\lambda_x^2v_x^2+\lambda_y^2v_y^2+\lambda_z^2 v_z^2)\nonumber\\
&+\frac{1}{2}\left(\frac{1}{v_x^2}+\frac{1}{v_y^2}+\frac{1}{v_z^2}\right)+\frac{K}{2(v_x v_y v_z)^2}.
\end{align}
To study the collective mode spectrum, Eq.~\eqref{eq:v} must be solved for the equilibrium widths $v_\eta$, and then linearized about harmonic perturbations to these equilibrium solutions $\delta v_\eta e^{-i\omega t}$.  The frequencies $\omega$ are eigenvalues of the $3\times3$ Hessian matrix $U_{ij}=\partial^2U/\partial_i\partial_j$
\begin{widetext}
\begin{align}
&\omega_{022}/\omega_\mathrm{ho}=\sqrt{U_{11}-U_{12}},\\
&\omega_{100}/\omega_\mathrm{ho}=\frac{1}{\sqrt{2}}\sqrt{U_{11}+U_{12}+U_{33}+\sqrt{U_{11}^2+2U_{11}U_{12}+U_{12}^2+8U_{13}^2-2U_{11}U_{33}-2U_{12}U_{33}+U^2_{33}}},\\
&\omega_{020}/\omega_\mathrm{ho}=\frac{1}{\sqrt{2}}\sqrt{U_{11}+U_{12}+U_{33}-\sqrt{U_{11}^2+2U_{11}U_{12}+U_{12}^2+8U_{13}^2-2U_{11}U_{33}-2U_{12}U_{33}+U^2_{33}}},
\end{align}
\end{widetext}
where the indices $\omega_{nlm}$ indicate the associated principle and angular quantum numbers for a particular mode \cite{PhysRevLett.77.5320}.  

Finally, we give analytic results from the variational calculations for a cylindrical trap ($\lambda_x=\lambda_y=1$).  In the limit $K\ll1$, we take $K=0$ to obtain the ground state widths $v_x=v_y=1$ and $v_{z}=1/\sqrt{\lambda_z}$, and find the spectrum
\begin{align}
&\omega_{022}/\omega_\mathrm{ho}=2,\\
&\omega_{100}/\omega_\mathrm{ho}=\begin{cases}
2\lambda_z\quad\quad& \text{if}~\lambda_z\geq1,\\
2\quad\quad &\text{if}~\lambda_z<1,\\
\end{cases}\\
&\omega_{020}/\omega_\mathrm{ho}=\begin{cases}
2\quad\quad& \text{if}~\lambda_z\geq1,\\
2\lambda_z\quad\quad &\text{if}~\lambda_z<1.\\
\end{cases}
\end{align}
In the limit $K\gg1$, we ignore the kinetic dispersion term to obtain the equilibrium widths $v=K^{1/8}\lambda_z^{1/4}$ and $v_{z}=K^{1/8}\lambda_z^{-3/4}$, which gives the spectrum
\begin{align}
&\omega_{022}/\omega_\mathrm{ho}=\sqrt{2},\\
&\omega_{100}/\omega_\mathrm{ho}=\sqrt{3+2\lambda_z^2+\sqrt{9-4\lambda_z^2+4\lambda_z^4}},\\
&\omega_{020}/\omega_\mathrm{ho}=\sqrt{3+2\lambda_z^2-\sqrt{9-4\lambda_z^2+4\lambda_z^4}}.
\end{align}

\end{document}